\RequirePackage{fix-cm}

\documentclass[twocolumn,epjc3]{svjour3}  
\smartqed  
\RequirePackage{graphicx}

\RequirePackage{float}
\RequirePackage{amsmath}
\RequirePackage{lineno}
\RequirePackage[colorlinks,citecolor=blue,urlcolor=blue,linkcolor=blue]{hyperref}
\journalname{Eur. Phys. J. C}
\modulolinenumbers[5] 

\begin{document}

\title{Puzzling time properties of proportional electroluminescence in two-phase argon detectors for dark matter searches
	}


\author{A.~Buzulutskov\thanksref{addr1,addr2}
        \and 
        E.~Frolov\thanksref{addr1,addr2,e1}
		\and
		E.~ Borisova\thanksref{addr1,addr2}
		\and
		V.~Nosov\thanksref{addr1,addr2}
		\and
		V.~Oleynikov\thanksref{addr1,addr2}
		\and
		A.~Sokolov\thanksref{addr1,addr2}
        }

\thankstext{e1}{geffdroid@gmail.com (corresponding author)}


\institute{Budker Institute of Nuclear Physics SB RAS, Lavrentiev avenue 11, 630090 Novosibirsk, Russia \label{addr1} 
\and Novosibirsk State University, Pirogova street 2, 630090 Novosibirsk, Russia \label{addr2}
}

\date{Received: date / Accepted: date}

\maketitle

\begin{abstract}
Proportional electroluminescence (EL) in noble gases is a physical process routinely used in two-phase (liquid-gas) detectors for low-energy astroparticle-physics experiments. In this work, the time properties of visible-light EL in two-phase argon detectors have been systematically studied for the first time.
In particular, two unusual slow components in the EL signal, with their contributions and time constants increasing with electric field, were observed. This puzzling property is not expected in any of the known mechanisms of photon and electron emission in two-phase media. Time constants of these components is about 4-5~$\mu$s and 50~$\mu$s.
In addition, a specific threshold behavior of the slow components was revealed: they emerged at a threshold in reduced electric field of 4.8~$\pm$~0.2~Td regardless of the gas phase density, which is about 1~Td above the onset of standard (excimer) EL.
There is a conspicuous similarity between this threshold and reduced field threshold of EL in NIR occurring via higher atomic excited states Ar$^{*}(3p^{5}4p)$.
An unexpected temperature dependence of slow components was also observed: their contribution decreased with temperature, practically disappearing at room temperature. We show that the puzzling properties of slow components can be explained in the framework of hypothesis that these are produced in the charge signal itself due to trapping of drifting electrons on metastable negative argon ions.
\end{abstract}


  \section{Introduction}\label{intro}
Proportional electroluminescence (EL) is used in two-phase dark matter detectors and low energy neutrino experiments~\cite{Chepel13,Buzulutskov20,Akimov21} to measure the primary ionization signal in the gas phase induced by particle scattering in the liquid phase. Such a signal (S2) is recorded with a delay relative to the primary scintillation signal (S1) which is recorded promptly. 
The current best limits on WIMP-nucleon spin-independent cross-section were obtained with two-phase detector configuration using both Xe~\cite{Aprile18,Aprile19} and Ar~\cite{Agnes18b,Agnes18} gases. In light of this success, there is interest both for more in-depth study of these detectors and for conducting next-generation experiments, in particular DarkSide-20k~\cite{Aalseth18}.

The study of time properties of the S2 signal, namely of its pulse shape, is of great importance for the correct interpretation of the data for dark matter searches, especially in the low-energy region where the ``S2 only'' analysis is applied~\cite{Agnes18,Aprile19}. In addition, the S2 pulse shape can be used to restore z-coordinate of the event~\cite{Aprile19,Agnes18a} and to measure the EL gap thickness to monitor potential sagging of the wire electrodes~\cite{Bondar20a}.

There is, however, some ambiguity regarding S2 pulse shapes in two-phase detectors: several abnormal slow components or delayed pulses were observed in Xe-based detectors on $\mu$s and ms scales~\cite{Aprile14,Akimov16,Sorensen18,Akerib20,Aprile21a,Kopec21}.
It was proposed that these delayed pulses were induced by delayed or trapped electrons, that is electrons which arrive at anode later than expected from drift time of primary ionization. The origin and quantitative description of these electrons is not yet determined and there are different interpretations and results across the references above. Nonetheless, these delayed electrons, following the main S2 signal, effectively increase single-electron noise of the detector. It was argued that this effect may explain excess of the S2-only signal rates compared to expected background which was observed earlier at low energies (low number of electrons) in the DarkSide-50~\cite{Agnes18} and XENON1T~\cite{Aprile20} experiments (see discussion in~\cite{Pereverzev22}).

On the other hand, no dedicated studies of slow components in EL (S2) signals have yet been carried out in Ar. In this work, we fill this gap: for the first time, the S2 time properties have been systematically studied in two-phase Ar detectors in a wide range of electric fields. The preliminary results of the study were presented in~\cite{Bondar20,Bondar20b}. In those works, two unusual slow components of S2 signals in two-phase Ar detector were observed, with time constants of about 4-5~$\mu$s and 50~$\mu$s, referred to as ``slow'' and ``long'' components respectively. The S2 signals were recorded both in the vacuum ultraviolet (VUV) range and in that of visible and near infrared (NIR). The spectral range of the latter is provided by the neutral bremsstrahlung (NBrS) EL mechanism, which was introduced in \cite{Buzulutskov18} and then further studied in \cite{Bondar20c,Takeda20,Kimura20,Aalseth21,Borisova21,Borisova22,Aoyama22,Amedo22,Henriques22}.

	\begin{figure}[!t]
	\centering
	\includegraphics[width=1.0\linewidth]{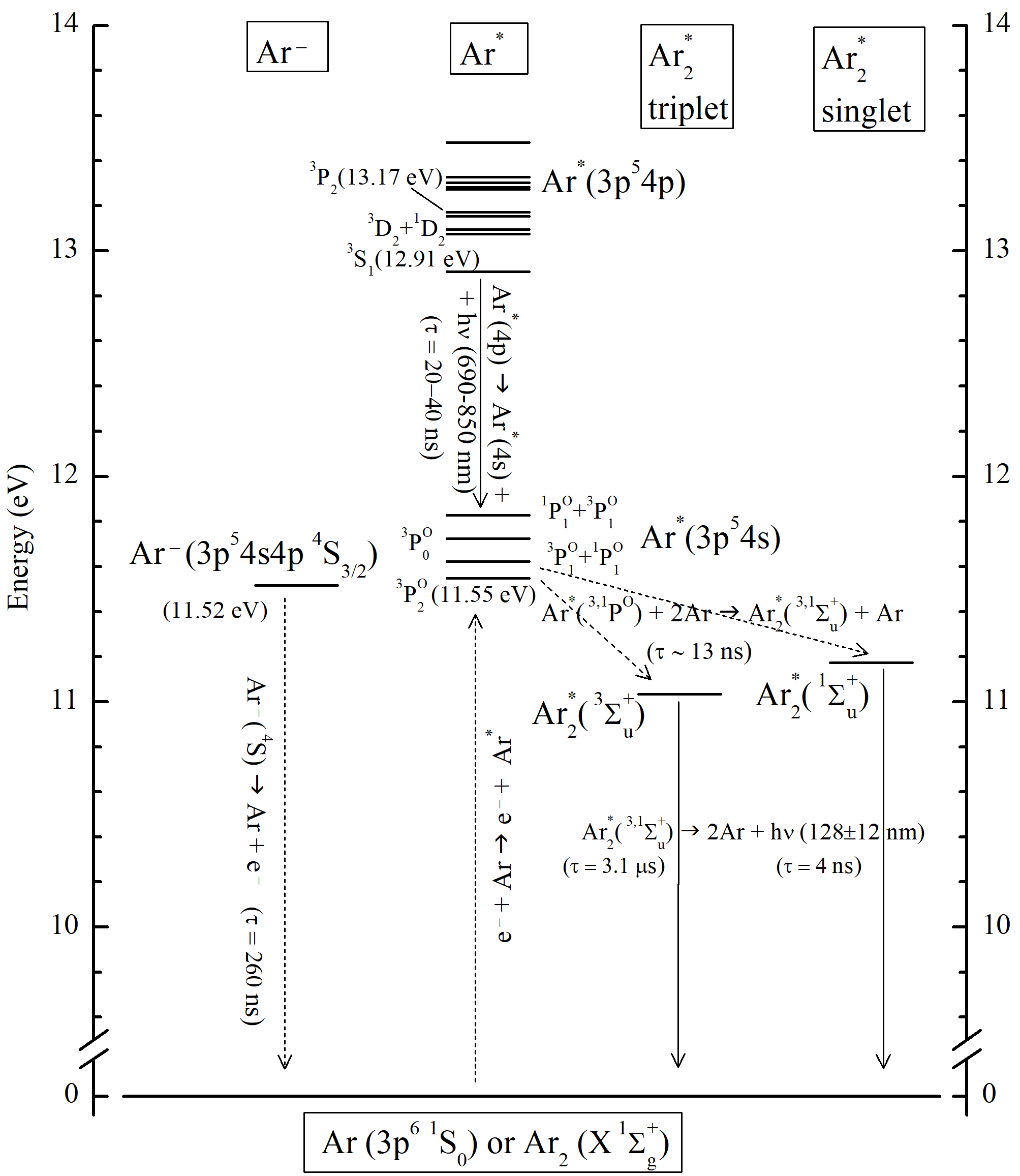}
	\caption{Energy levels of atomic excited states Ar$^*$ \cite{NIST,Buzulutskov17}, molecular excited states Ar$_2^{\ast}$ (minimums of potential energy curves~\cite{Yates83,Duplaa96}) and metastable negative ion state Ar$^{-}$ \cite{Bae1985:Ar_ion_state,Ben-Itzhak1988:Ar_ion_state}, all given in LS-coupling notation. The atomic levels are shown with their mixing determined from~\cite{Katsonis11}. The solid arrows indicate the radiative transitions observed in experiments: Ar$_2^{\ast}$ in the VUV \cite{Cheshnovsky72} and Ar$^{\ast}$ in the NIR \cite{Lindblom88,Hofmann13}. Time constants for the VUV and NIR transitions are taken from~\cite{Buzulutskov17} and~\cite{Schulze08} respectively. The dashed arrows indicate the non-radiative transitions. The ground states are composed of argon atoms $\mathrm{Ar}\,(3p^{6}\, ^1S_0)$ and Van-der-Waals dimers $\mathrm{Ar}_2(\mathrm{X}\, ^1\Sigma^{+}_{g})$.}
	\vspace{-10pt}
	\label{fig:Ar_levels}
\end{figure}

According to modern concepts~\cite{Buzulutskov20}, there are three mechanisms responsible for proportional EL in noble gases: that of excimer (Ar$^*_2$) emission in the VUV \cite{Oliveira11}, that of emission due to atomic transitions in the NIR \cite{Oliveira13,Buzulutskov17}, and that of NBrS emission in the UV, visible and NIR range~\cite{Buzulutskov18}. These three mechanisms are referred to as excimer (ordinary) EL, atomic EL and NBrS EL, respectively. The first two are illustrated in Fig.~\ref{fig:Ar_levels}.

Neutral bremsstrahlung (NBrS) EL is due to bremsstrahlung of drifting electrons elastically scattered on neutral atoms \cite{Buzulutskov18,Borisova21,Henriques22}:
\begin{eqnarray}
\label{Rea-NBrS-el}
e^- + \mathrm{Ar} \rightarrow e^- + \mathrm{Ar} + h\nu \; . 
\end{eqnarray}
It has no threshold in energy and thus in electric field. This process is extremely fast, with characteristic time less than 1~ps.

Excimer EL is due to emission of noble gas excimers, in a singlet (Ar$^{*}_{2}(^{1}\Sigma^{+}_{u})$) or triplet (Ar$^{*}_{2}(^{3}\Sigma^{+}_{u})$) state, produced in three-body atomic collisions of the lowest excited atomic states, of Ar$^*(3p^54s)$ configuration, which in turn are produced by drifting electrons in electron-atom collisions (see reviews~\cite{Chepel13,Buzulutskov20,Buzulutskov17}):
\begin{eqnarray}
\label{Rea-ord-el}
e^- + \mathrm{Ar} \rightarrow e^- + \mathrm{Ar}^{\ast}(3p^54s) \; , \nonumber \\
\mathrm{Ar}^{\ast}(3p^54s) + 2\mathrm{Ar} \rightarrow \mathrm{Ar}^{\ast}_2(^{1,3}\Sigma^{+}_{u}) + \mathrm{Ar} \; , \nonumber \\
\mathrm{Ar}^{\ast}_2(^{1,3}\Sigma^{+}_{u}) \rightarrow 2\mathrm{Ar} + h\nu \; .
\end{eqnarray}

Atomic EL is due to atomic transitions between the higher (Ar$^*(3p^54p)$) and lower (Ar$^*(3p^54s)$) excited states, the former being also produced by drifting electrons~\cite{Buzulutskov11,Oliveira13}:
\begin{eqnarray}
e^- + \mathrm{Ar} \rightarrow e^- + \mathrm{Ar}^{\ast}(3p^54p) \; , \nonumber \\
\mathrm{Ar}^*(3p^54p) \rightarrow \mathrm{Ar}^*(3p^54s)+h\nu \; .
\end{eqnarray}

All mentioned EL mechanisms are described using energy distribution of drifting electrons which is dependent on reduced electric field, $E/N$, where $E$ is the electric field and $N$ is the atomic density. For this reason, in the figures presented in this work, many quantities are shown as a function of the reduced electric field. It is expressed in Td units: 1~Td~=~$10^{-17}$~V~cm$^2$, corresponding to the electric field of 0.87~kV/cm in gaseous Ar in the two-phase mode at 87.3~K and 1.00 atm.

The puzzling property of the observed slow components is that their contributions and time constants increase with electric field. Such an increase does not correspond to either of the two known mechanisms of slow component formation in two-phase Ar detectors, namely to that of thermionic electron emission from liquid to gas phase~\cite{Borghesani90,Bondar09} and that of VUV scintillation via triplet excimer state Ar$_{2}^{*}$($^{3}\Sigma_{u}^{+}$)~\cite{Buzulutskov17}. In the first mechanism, slow component is caused by electrons which are delayed on potential barrier at liquid-gas interface. The higher electric field, the easier it is for electrons to satisfy conditions to exit the liquid, which results in decrease of both contribution and time constant of slow component with increasing field. In particular, their values in argon are less than about 4\% and 2~$\mu$s respectively for electric field in liquid of 2.5~kV/cm~\cite{Borghesani90,Bondar09}.

In the case of excimer scintillation in VUV, there are no delayed electrons. Rather, there is delay in emitting photon which is caused by quite long (3.1~$\pm$~0.1~$\mu$s) lifetime of triplet state Ar$_{2}^{*}$($^{3}\Sigma_{u}^{+}$). If detector is made sensitive to VUV, then recorded EL signal is defined almost fully by excimer scintillation mechanism for the typical electric fields. In this case, the fast and slow components of EL signal are produced by the singlet (4.2~ns lifetime) and by the triplet states respectively. The time constant of excimer slow component is equal to the lifetime of triplet state and its contribution is defined by ratio of triplet and singlet states formed in three-body collisions (see Eq.~\ref{Rea-ord-el}). Thus, both time constant and contribution of excimer slow component should not depend on the electric field under condition that VUV scintillation dominates recorded EL signal.

In our preliminary work~\cite{Bondar20} we studied the unusual slow components using gamma-rays in detector configuration with and without wavelength shifter (tetraphenyl butadiene (TPB)), that is in configuration sensitive to excimer scintillation in VUV and sensitive only to the scintillation in visible and NIR range respectively. In the preliminary work, it was supposed that these components are most likely present in the charge signal itself and that metastable negative Ar ions might be responsible for their formation. In this work, we study the unusual slow components in a more elaborated way, in particular using an alpha-particle source in addition to that of gamma-ray. Moreover, to clarify the issue of the origin of unusual slow components, we study their new properties: the threshold behavior with respect to electric field, the dependence on pressure (i.e. on density of the gas phase) and the temperature dependence.

We show in section~\ref{sec:hypotheis} that the results of this study strongly support the hypothesis of metastable negative argon ions being responsible for the formation of slow components. Such negative ion states can be produced either in 3-body collisions of drifting electrons with Ar atoms or collisions with Van-der-Waals molecules $\mathrm{Ar}_2(\mathrm{X}\, ^1\Sigma^{+}_{g})$~\cite{Buzulutskov17,Smirnov84,Smirnov96,Stogrin59}.

If our hypothesis is correct, then similarly to Xe, Ar detectors have delayed electrons correlated to S2 signal. This will effectively increase single-electron noise of the detector, which in turn will affect expected rates of events with small number of electrons. Thus, this effect may change the sensitivity of Ar detectors at low dark matter masses, near detector threshold.

  \section{Experimental setup}\label{SetupSection}

	\begin{figure}[!t]
	\centering
	\includegraphics[width=1.0\linewidth]{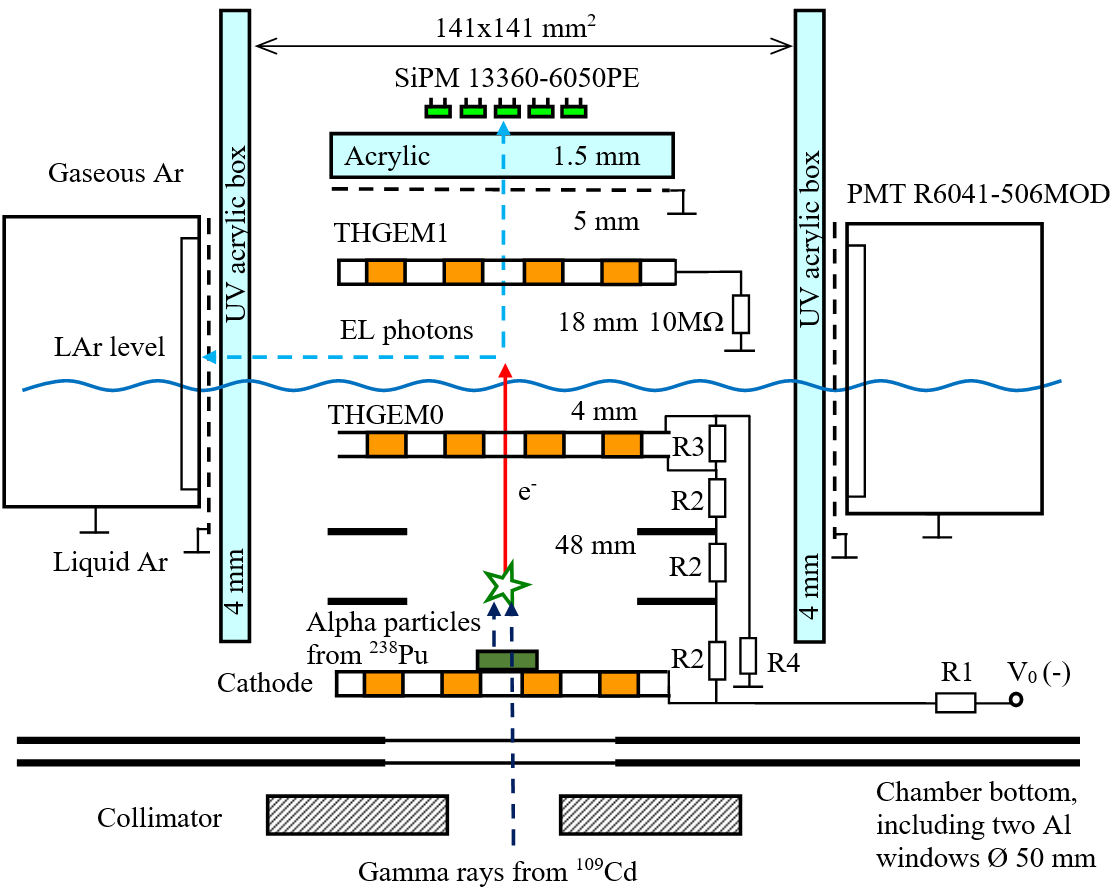}
	\caption{Schematic view of the experimental setup (not to scale). Dimensions of drift region (48~mm), electron emission region (4~mm) and EL gap (18~mm) are also shown.}
	\vspace{-10pt}
	\label{fig:setup_scheme}
	\end{figure}

	\begin{figure}[!t]
	\centering
	\includegraphics[width=1.0\linewidth]{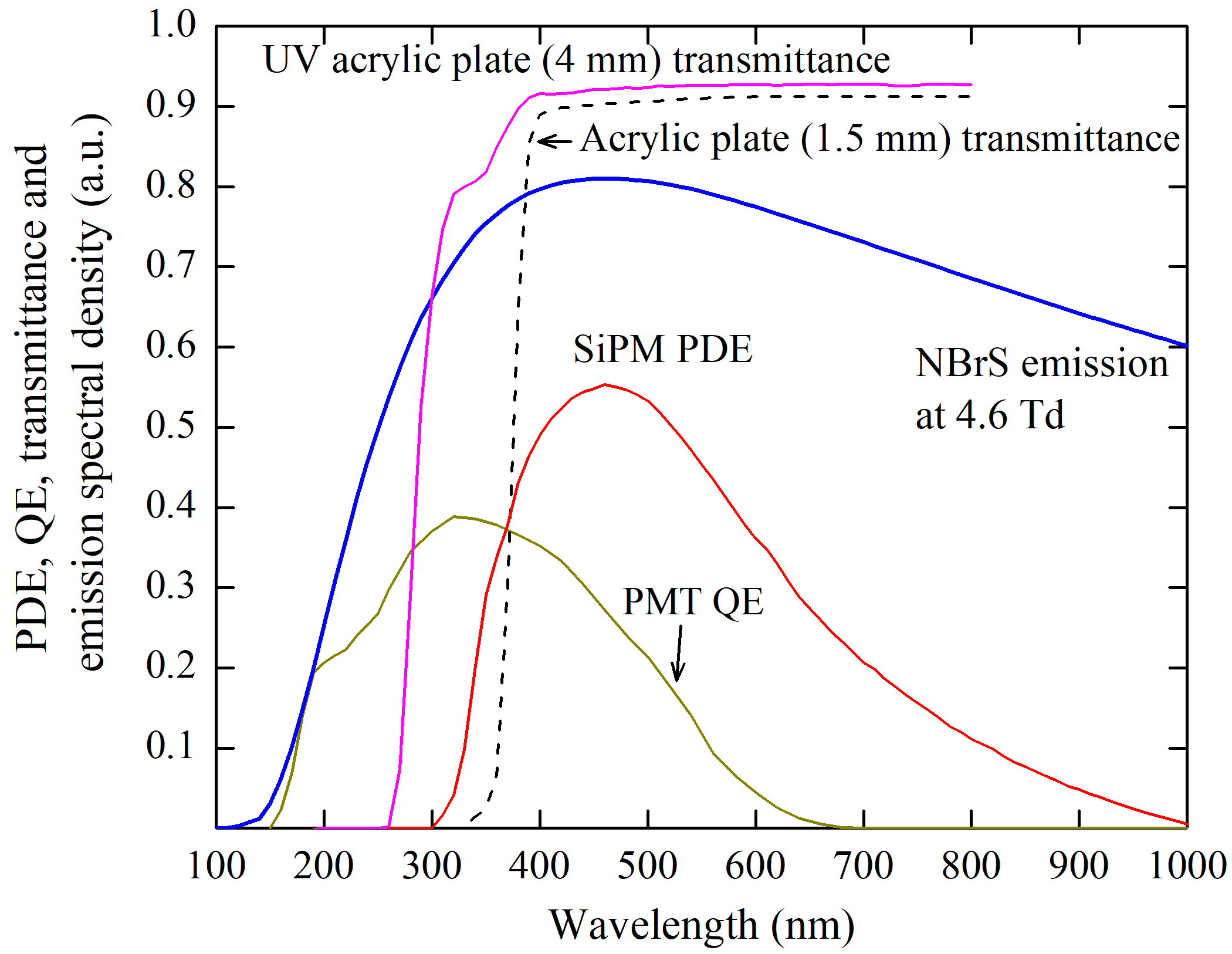}
	\caption{Quantum efficiency (QE) of the PMT R6041-506MOD at 87~K obtained from~\cite{Hamamatsu,Lyashenko14}, Photon Detection Efficiency (PDE) of the SiPM (MPPC 13360-6050PE~\cite{Hamamatsu}) at an 5.6~V overvoltage obtained from~\cite{Otte17}, transmittance of the ordinary and UV acrylic plate in front of the SiPM and bare PMT respectively, measured by us using light source, monochromator and calibrated photodiode. Also shown is the emission spectrum of neutral bremsstrahlung electroluminescence (NBrS EL) at reduced electric field of 4.6~Td~\cite{Buzulutskov18,Borisova21}.}
	\vspace{-10pt}
	\label{fig:setup_spectra}
	\end{figure}

A schematic view of the two-phase detector used in this work is shown in Fig.~\ref{fig:setup_scheme}. The detector included a 9~L cryogenic  chamber filled with 2.5-3.5~liters of liquid Ar and operated in a two-phase mode in equilibrium state at a saturated vapor pressure of 1.00, 1.50 or 0.75~atm, corresponding to a temperature of 87.3, 91.3 or 84.7~K respectively~\cite{Fastovsky72}. In addition to usual working pressure of 1.00~atm, the lowest and the highest possible values were used, the former being limited by argon triple point and cryogenics performance, and the latter being limited by durability of Al windows at the bottom of the chamber (see Fig.~\ref{fig:setup_scheme}). Compared to the detector used in preliminary studies \cite{Bondar20,Bondar20b}, there were only minor modifications. In particular, the detector configuration without wavelength shifter (WLS) was used, i.e. with direct optical readout in the visible and NIR range using PMTs and a SiPM matrix. 

Thick gas electron multiplier (THGEM) electrodes of 10$\times$10~cm$^2$ active area were used to form drift region (between the cathode and THGEM0), electron emission region (above THGEM0) and EL gap (between the liquid surface and THGEM1 anode). These regions had electric fields of 0.093-0.68~kV/cm, 0.71-5.2~kV/cm and 1.1-8.0~kV/cm respectively, the voltage applied to the divider varying from 3 to 22~kV.
Electric field uniformity of this geometry was studied in~\cite{Bondar19} and found to be satisfactory, meaning that expected systematic error from non-uniformity was determined to be much less than other systematic errors of values we measure.
During the measurements, the EL gap thickness could be decreased by liquefying an additional amount of Ar.
In particular, two values of 18 and 10~mm were used in this work.
The liquid level was calculated from the amount of condensed Ar using CAD software and was verified in calibration runs using THGEM1 as a capacitive liquid level meter. 

At the start of each experimental run, Ar gas was liquefied from a storage bottle into the cryogenic chamber while passing through Oxysorb filter in order to purify it from electronegative impurities. The initial Ar total impurity content declared by manufacturer was below 2~ppm.
This filtration process was demonstrated to achieve electron life-time in liquid Ar $>$~100~$\mu$s at 200~V/cm field~\cite{Bondar17}, which corresponds to oxygen content below few ppb.
Additionally, the N$_2$ content was monitored by a gas analyzer SVET~\cite{SVET} based on an emission spectrum measurement technique: it was below 1~ppm. At the end of the run, Ar was collected from the chamber back into the bottle cooled with liquid nitrogen. 

In this work, the S2 signals from the EL gap were optically recorded in the visible and NIR range using the effect of NBrS EL \cite{Buzulutskov18,Aalseth21}. Optical readout of the EL gap was provided by four compact PMTs R6041-506MOD \cite{Bondar15,Bondar17a}, placed on every side of the gap, and by a 5$\times$5 SiPM matrix composed of 13360-6050PE type SiPMs~\cite{Hamamatsu}, with 1 cm pitch, located above the anode. Their quantum efficiency (QE) and photon detection efficiency (PDE), respectively, are shown in Fig.~\ref{fig:setup_spectra}, along with example of the NBrS EL spectrum. Each PMT and SiPM channel was read out individually at 62.5~MS/s rate for a total waveform of 160~$\mu$s duration. The sum of all the PMT channels was used for triggering by the S2 signal. Even though the SiPM matrix detects larger number of photoelectrons, correlated noise prevented us from using sum of even few of its channels for triggering.

To reach the PMTs, the photons produced in the EL gap passed through 4 mm thick walls of the UV acrylic box, while to reach the SiPMs the photons passed through a 1.5~mm thick acrylic plate (see Fig.~\ref{fig:setup_spectra} for their transmission spectra). In addition, the THGEM1 plate in front of the SiPM matrix acted as an optical mask; it had dielectric thickness of 0.94~mm, hole pitch of 1.1 mm, hole diameter of 1.0 mm, and thus provided a 75\% optical transparency at normal incidence. 

\begin{figure*}[th!]
	\centering
	\includegraphics[width=1.0\linewidth]{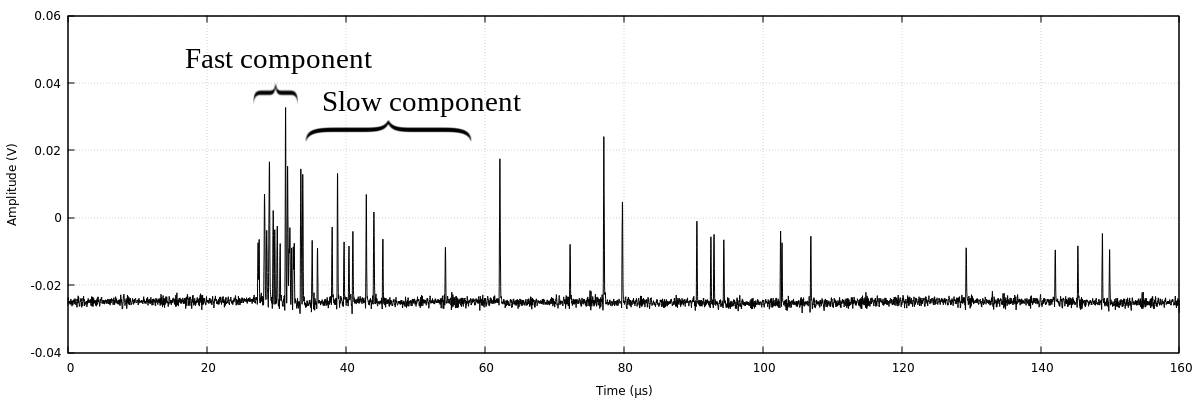}
	\caption{Example of S2 signal event (waveform) from one of the SiPM channels obtained with $^{238}$Pu alpha particles due to NBrS EL in 18~mm thick EL gap, at reduced electric field of 8.0~Td and pressure of 1.00~atm.} 
	\vspace{-10pt}
	\label{fig:signal_example}
\end{figure*}

\begin{figure}[!tb]
	\centering
	\includegraphics[width=1.0\linewidth]{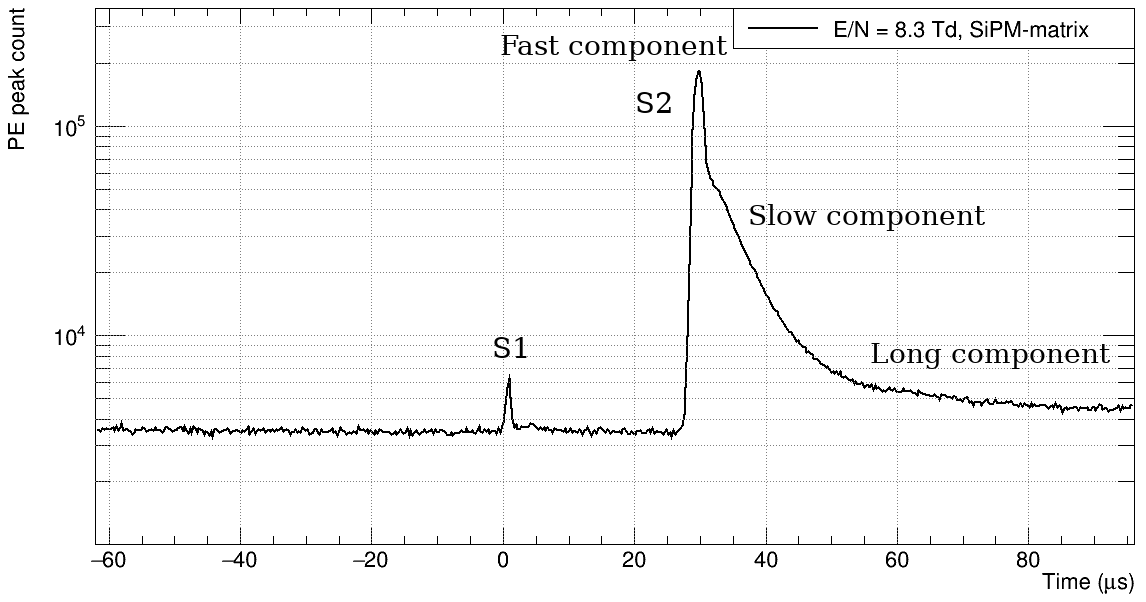}
	\caption{Raw pulse shape for the given experimental run. Shown is the sum of signal waveforms over 25 channels of SiPM matrix and selected events obtained in two-phase Ar detector with $^{238}$Pu alpha particles due to NBrS EL in 18~mm thick EL gap, at reduced electric field of 8.3~Td and pressure of 1.00~atm. The trigger was provided by the S2 signal.}
	\vspace{-10pt}
	\label{fig:slow_comp_alpha_log}
\end{figure}

\begin{figure}[!tb]
	\centering
	\includegraphics[width=1.0\linewidth]{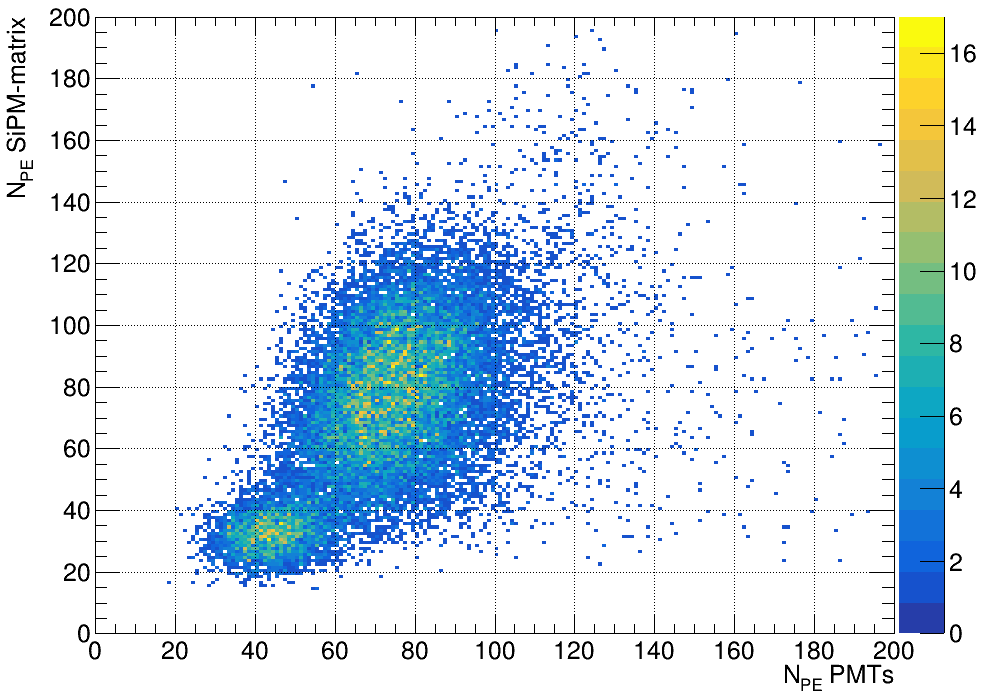}
	\caption{2D distribution of SiPM-matrix amplitude versus 4 PMT amplitude, both expressed in the number of photoelectrons (N$_{PE}$), obtained in two-phase Ar detector with $^{109}$Cd gamma rays due to NBrS EL in 18~mm thick EL gap, at reduced electric field of 8.3~Td and pressure of 1.00~atm. One can see high-energy area composed of 58-69 and 88 keV lines and that of low-energy composed of 22-25 keV lines, as well a cutoff at low energies (at about 20 photoelectrons) defined by the PMT trigger threshold.} 
	\vspace{-10pt}
	\label{fig:Npe_SiPM_PMT}
\end{figure}

\begin{figure}[!tb]
	\centering
	\includegraphics[width=1.0\linewidth]{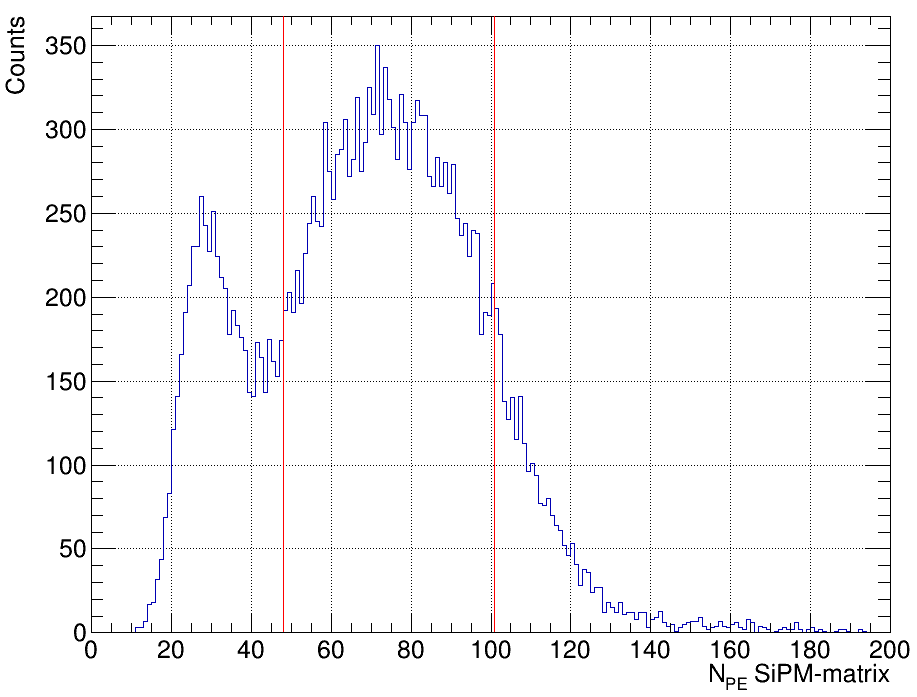}
	\caption{Distribution of SiPM-matrix amplitudes expressed in the number of photoelectrons, obtained in two-phase Ar detector with $^{109}$Cd gamma rays due to NBrS EL in 18~mm thick EL gap, at reduced electric field of 8.3~Td and pressure of 1.00~atm. This figure is the projection of Fig.~\ref{fig:Npe_SiPM_PMT} on vertical axis.} 
	\vspace{-10pt}
	\label{fig:Npe_SiPM_Cd}
\end{figure}

\begin{figure}[!tb]
	\centering
	\includegraphics[width=1.0\linewidth]{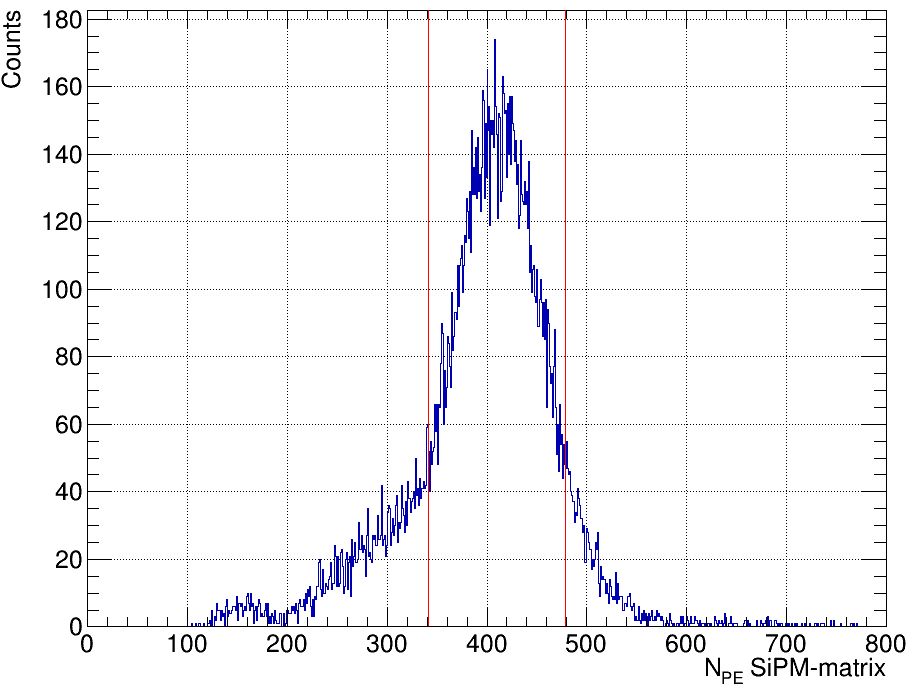}
	\caption{Distribution of SiPM-matrix amplitudes expressed in the number of photoelectrons, obtained in two-phase Ar detector with $^{238}$Pu alpha particles due to NBrS EL in 18~mm thick EL gap, at reduced electric field of 8.0~Td and pressure of 1.00~atm.} 
	\vspace{-10pt}
	\label{fig:Npe_SiPM_Pu}
\end{figure}

\begin{figure}[!tb]
	\centering
	\includegraphics[width=1.0\linewidth]{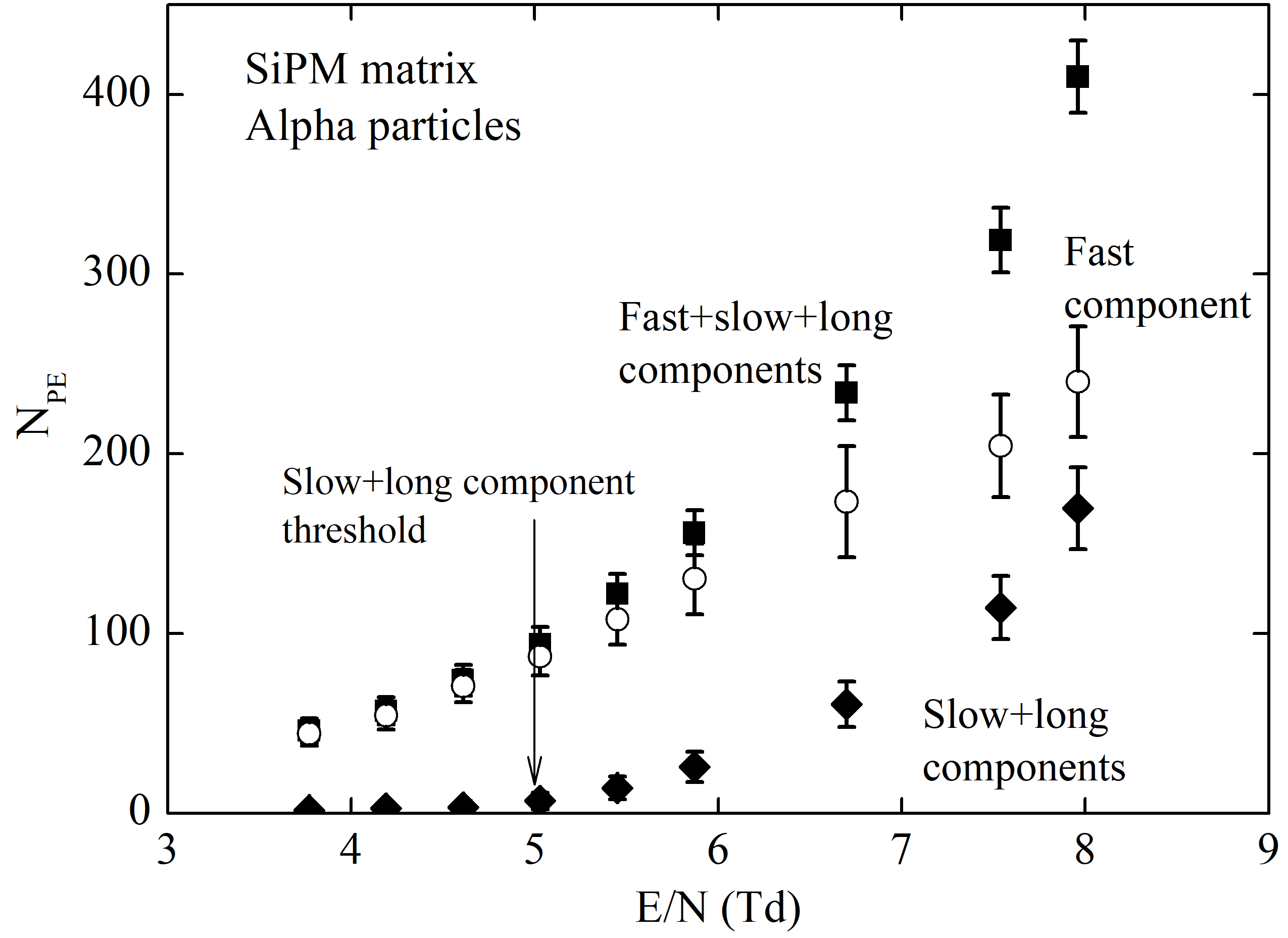}
	\caption{Amplitude of the total S2 signal (fast+slow+long components)  and its time components separately (fast and slow+long components), expressed in the number of photoelectrons recorded by the SiPM matrix, as a function of the reduced electric field, obtained in two-phase Ar detector with $^{238}$Pu alpha particles due to NBrS EL in 18~mm thick EL gap at 1.00~atm. The arrow points to the threshold of slow components appearance.}
	\vspace{-10pt}
	\label{fig:Alpha_Npe_E}
\end{figure}

Two radioactive sources were used for S2 pulse-shape measurements: a $^{109}$Cd gamma-ray source on W substrate, having high-energy (58-69 and 88 keV) and low-energy (22-25 keV) lines~\cite{Bondar19a}, and a $^{238}$Pu alpha-particle source having a 5.5~MeV line. The latter was placed on the top of the cathode (Fig.~\ref{fig:setup_scheme}). When $^{238}$Pu was used in the detector, the gamma rays from $^{109}$Cd source did not reach active volume due to absorption in Pu substrate (2~mm of steel). Thus each experimental run was conducted only using one of the two radioactive sources. 

In measurements of the temperature dependence of slow components contribution, the detector configuration with alpha-particle source was used. In order to change temperature while having fixed Ar density, it is necessary to use single-phase mode (only gaseous Ar), since the two-phase mode dictates that temperature and density are directly linked according to saturated vapor curve. The measurements were conducted at different temperatures and at the same Ar gas atomic density, of 3.7$\cdot 10^{19}$~cm$^{-3}$, corresponding to 1.5~atm pressure at room temperature, the chamber temperature being varied from that of room (295~K) to 120~K. The mean path length of 5.5~MeV alpha particles at this gas density is 3.1~cm which corresponds up to 11.7~$\mu$s drift time at 0.24~kV/cm (0.63~Td) if the track is collinear to the electric field. For this reason, the tracks orthogonal to the electric field had to be selected in order to localize ionization along z-axis. Otherwise, the resulting EL pulse shapes are complicated by interplay of non-trivial charge distribution and unusual slow components.

More detailed description of the detector, trigger, as well as full procedure of data analysis can be found in our preliminary work~\cite{Bondar20}.

\section{Results}
\subsection{Measurements of S2 amplitudes and pulse shapes in two-phase detector}\label{results_general}

As was demonstrated in \cite{Bondar20c,Aalseth21}, the NBrS effect allows for direct recording the EL (S2) signals in two-phase Ar detectors, i.e. without using WLS. An example waveform of such S2 signal event obtained with alpha-particle source is shown in Fig.~\ref{fig:signal_example}. Each peak in the waveform corresponds to at least one photoelectron (PE) produced by a photon in the SiPM. This figure clearly demonstrates the presence of both the fast and slow components in the signal.

In addition, Fig.~\ref{fig:slow_comp_alpha_log} shows the sum of such waveforms over all the 25 channels of the SiPM matrix and over all selected events, reflecting the raw pulse shape for the given experimental run. In addition to fast and slow components, the latter having a time constant of about 5~$\mu$s, one can also see the second slow component, further referred to as ``long'' component, with a tenfold larger time constant, of about 50~$\mu$s. The weak primary scintillation (S1) signal was also observed at about 30~$\mu$s before the S2 signal (for its details see the recent study on visible-light scintillations in liquid Ar~\cite{Bondar22}).

For the correct analysis of the S2 pulse shape it is necessary to separate the events induced by radioactive sources from background (mostly caused by cosmic rays and radioactivity of materials inside the detector). This selection was provided by the analysis of the S2 amplitudes expressed in the number of photoelectrons (N$_{PE}$) recorded by the SiPM matrix and PMTs: see Figs.~\ref{fig:Npe_SiPM_PMT},~\ref{fig:Npe_SiPM_Cd} and~\ref{fig:Npe_SiPM_Pu}. One can see that for the $^{109}$Cd gamma-ray source the high-energy  peak, composed of 58-69 and 88~keV lines, is well separated from that of low energy, composed of 22-25 keV lines. In these figures only single events are shown, double and other multiple superimposed events being discarded (see~\cite{Bondar20} for details).

The red lines in Figs.~\ref{fig:Npe_SiPM_Cd} and~\ref{fig:Npe_SiPM_Pu} show amplitude cuts for selecting events for further analysis of the S2 pulse shape, namely the events of high-energy peaks of the $^{109}$Cd and $^{238}$Pu sources. Note that the S2 amplitude of high-energy peak for Pu source (420~PE) is six times larger than that of Cd source (70~PE). High-energy peaks were selected for several reasons. Firstly, they provide larger statistics for the S2 pulse-shapes due to larger number of PE. Secondly, larger number of PE means better selection of single events (see detailed procedure in~\cite{Bondar20}) which is especially crucial at low electric fields. Lastly, low-energy peaks have larger fraction of background events. 

Fig.~\ref{fig:Alpha_Npe_E} shows the electric field dependence of the amplitude of the total S2 signal, which includes all time components. A noticeable amplitude of the signal below the threshold of excimer EL, i.e. below 4 Td, reflects the NBrS nature of proportional EL (see section~\ref{intro} for details). The fast rise of the amplitude with electric field is explained by two effects: by field dependence of ionization yield from alpha-particle tracks in liquid Ar~\cite{Anderson88} and by field dependence of NBrS EL itself~\cite{Buzulutskov20,Borisova21}. 

To study the time properties of EL signals in detail, the time histogram of the recorded photoelectrons for the selected events was produced (see Fig.~\ref{fig:shapes_composition}). Each photoelectron gave one entry in the histogram with its own recorded time. The resulting histogram thus reflects the real S2 pulse shape averaged over all selected events. Only such signal shapes for the selected $^{109}$Cd and $^{238}$Pu events are used and analyzed.

\subsection{Electric field dependence and threshold behavior of slow components}\label{results_threshold}

	\begin{figure*}[!t]
	\centering
	\includegraphics[width=1.0\linewidth]{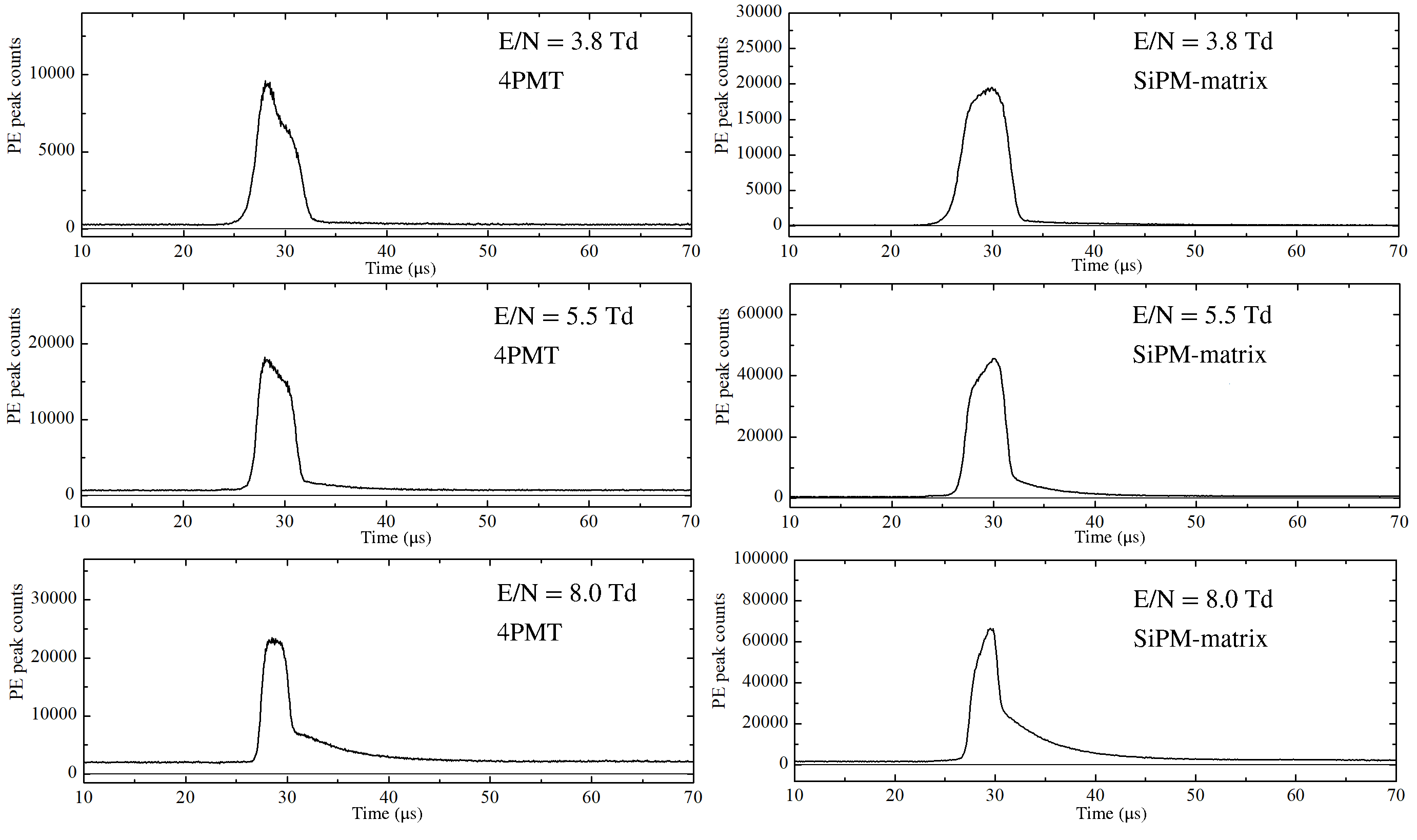}
	\caption{S2 pulse shapes in two-phase Ar detector for the 4PMT
(left) and SiPM-matrix (right) readout at low (3.8 Td), middle (5.5 Td) and high (8.0 Td) reduced electric field, obtained with $^{238}$Pu alpha particles due to NBrS EL in 18~mm thick EL gap at 1.00~atm.}
	\vspace{-10pt}
	\label{fig:shapes_composition}
	\end{figure*}

Fig.~\ref{fig:shapes_composition} illustrates how the slow components emerge and how much they contribute to the overall signal. The S2 pulse shapes are shown at low (3.8 Td), middle (5.5 Td) and high (8.0 Td) reduced electric fields. At low field, as expected, only the fast component due to NBrS EL is present in the signal. Its shape is defined mainly by the electron drift time across the EL gap, and, to a lesser extent, by electron diffusion, trigger conditions and light collection non-uniformity~\cite{Agnes18a,Bondar20}.

\begin{figure}[!t]
	\centering
	\includegraphics[width=1.0\linewidth]{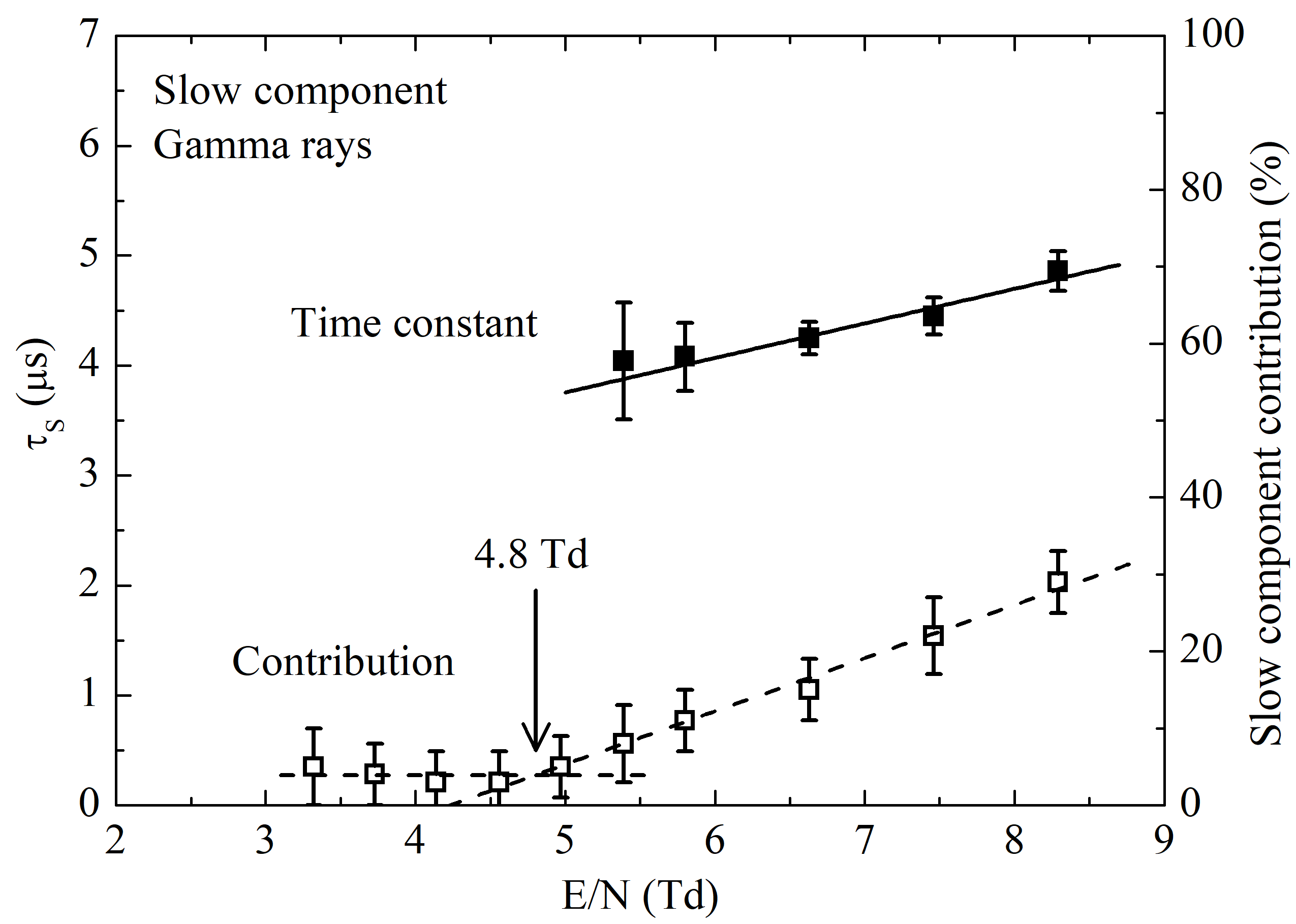}
	\caption{Time constant ($\tau_{S}$) and contribution to overall signal of the slow component as a function of the reduced electric field, obtained in two-phase Ar detector with $^{109}$Cd gamma rays due to NBrS EL in 18~mm thick EL gap at 1.00~atm. The arrow points to the nominal threshold of slow component appearance. The PMT and SiPM data are averaged.}
	\vspace{-10pt}
	\label{fig:slow_comp_gamma}
\end{figure}

\begin{figure}[!t]
	\centering
	\includegraphics[width=1.0\linewidth]{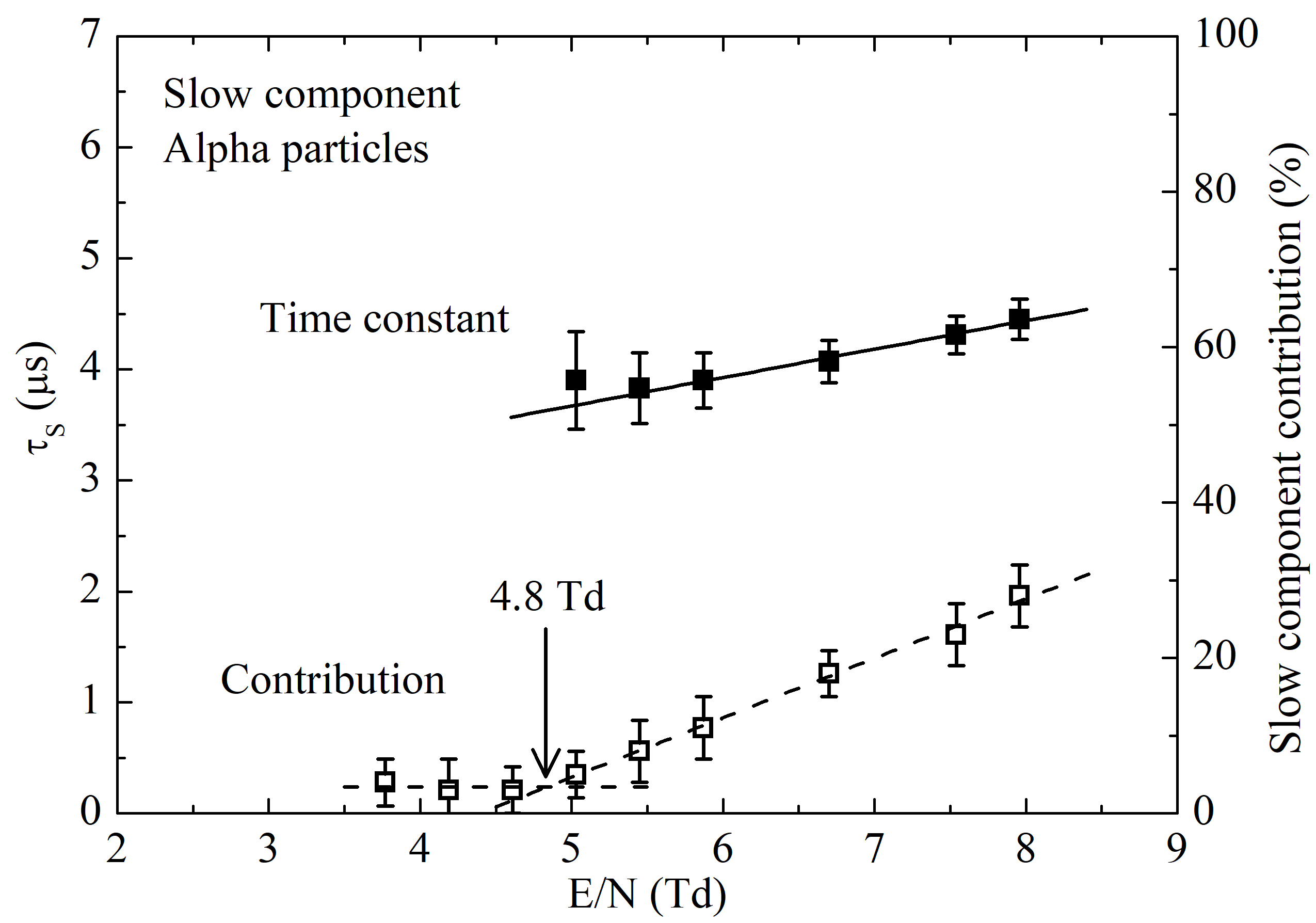}
	\caption{Time constant ($\tau_{S}$) and contribution to overall signal of the slow component as a function of the reduced electric field, obtained in two-phase Ar detector with $^{238}$Pu alpha particles due to NBrS EL in 18~mm thick EL gap at 1.00~atm. The arrow points to the nominal threshold of slow component appearance. The PMT and SiPM data are averaged.}
	\vspace{-10pt}
	\label{fig:slow_comp_alpha}
	\end{figure}

\begin{figure}[!t]
	\centering
	\includegraphics[width=1.0\linewidth]{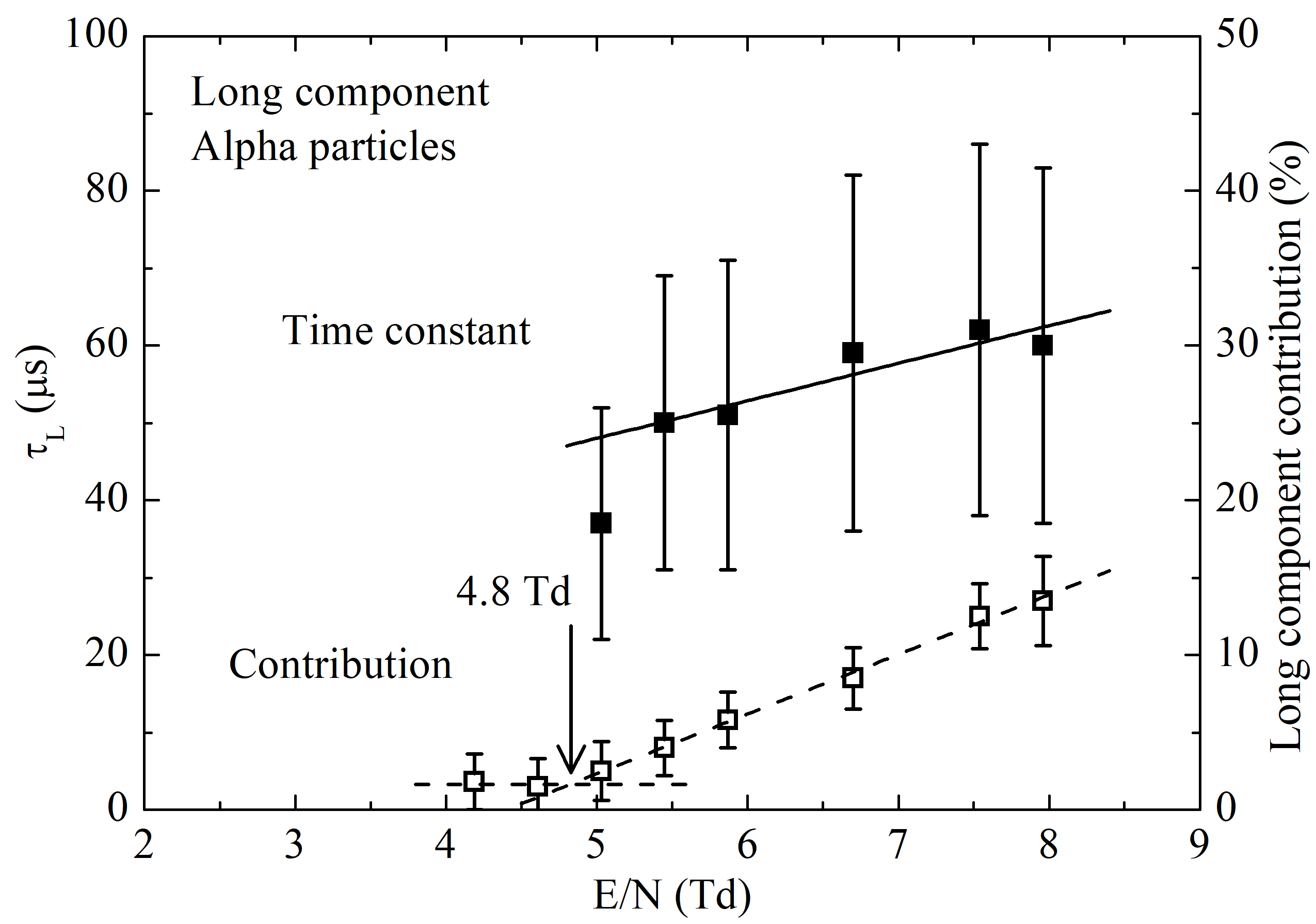}
	\caption{Time constant ($\tau_{L}$) and contribution to overall signal of the long component as a function of the reduced electric field, obtained in two-phase Ar detector with $^{238}$Pu alpha particles due to NBrS EL in 18~mm thick EL gap at 1.00~atm. The arrow points to the nominal threshold of long component appearance. The PMT and SiPM data are averaged.}
	\vspace{-10pt}
	\label{fig:long_comp_alpha}
	\end{figure}

At higher fields, exceeding 5~Td, a 4-5~$\mu$s slow component emerges and increases with electric field in both contribution and time constant: see Figs.~\ref{fig:Alpha_Npe_E},\ref{fig:slow_comp_gamma}, \ref{fig:slow_comp_alpha}, \ref{fig:long_comp_alpha}. A 50~$\mu$s long component also becomes apparent on logarithmic scale, as seen in Fig.~\ref{fig:slow_comp_alpha_log}, with similar electric field dependence. At reduced electric fields exceeding 8.0 Td the total contribution of the slow and long components to the overall signal becomes quite large, reaching 50\%. One can also see from Fig.~\ref{fig:shapes_composition} that the PMT and SiPM-matrix data are in good agreement; accordingly, all further results show time constants and contributions as average of fitted values for the PMTs and the SiPM-matrix. 

It should be noted that the contributions of the fast and slow components are defined here in the same way as in~\cite{Bondar20}. Namely, the pulse area before the time point of the characteristic bend at around 32~$\mu$s (see Fig.~\ref{fig:shapes_composition}) is considered to be fully fast component and everything after is considered to be only the slow and long components. The slow and long components are separated from each other using exponential fits. Proper separation from the fast component requires a specific physical model of the slow components, but the current "rough" approach is sufficient for our purposes.

The increase of contribution and time constants of slow and long components with electric field can not be explained by either known mechanism of slow component formation in EL signal of two-phase Ar detectors: neither by electron emission from liquid to gas phase~\cite{Borghesani90,Bondar09}, nor by photon emission via Ar$_{2}^{*}$($^{3}\Sigma_{u}^{+}$) triplet excimer state in the VUV~\cite{Buzulutskov17}. Indeed, in the excimer emission mechanism, providing a 3.1-3.3~$\mu$s slow component due to the triplet excimer state Ar$^{*}_{2}(^{3}\Sigma^{+}_{u})$, the emission time constant and contribution (relative to the fast component provided by singlet excimer state Ar$^{*}_{2}(^{1}\Sigma^{+}_{u})$) do not depend on the electric field~\cite{Buzulutskov17}. In the mechanism of electron emission from liquid into gas phase, responsible for the slow component in S2 signal at lower extraction electric fields (below 2 kV/cm), the slow component time constant and contribution decrease with electric field~\cite{Borghesani90,Bondar09}. Hence we call the slow components in Figs.~\ref{fig:slow_comp_alpha_log} and \ref{fig:shapes_composition} ``unusual''.

To better understand the nature of these components, their threshold behavior has been investigated. The results are presented in Fig.~\ref{fig:Alpha_Npe_E} in terms of the amplitude characteristics and in Figs.~\ref{fig:slow_comp_gamma}, \ref{fig:slow_comp_alpha} and \ref{fig:long_comp_alpha} in terms of the slow components contribution to overall signal. These figures clearly demonstrate that the amplitude and contribution of the slow and long components has a certain threshold in the reduced electric field. In Figs.~\ref{fig:slow_comp_gamma}, \ref{fig:slow_comp_alpha} and \ref{fig:long_comp_alpha} the nominal threshold is defined as the intersection of a linearly increasing function fitted to data above the threshold and a horizontal line fitted to data below the threshold, the latter corresponding to background baseline.

Total systematic uncertainty of threshold value was estimated to be about 0.2~Td (standard deviation). This uncertainty has two main contributions of about the same magnitude. The first contribution is an uncertainty of electric fields values ($\approx$0.12~Td) due to an uncertainty of the EL gap thickness (about 1~mm). The second one ($\approx$0.14~Td) is due to uncertainty of fitted lines determining the nominal threshold both due to inclusion or exclusion of points near the threshold from the fits and due to errors of the points themselves. The main error of the latter is systematic error due to event selection.

It should be noted that large errors of long component time constant in Fig.~\ref{fig:long_comp_alpha} are also almost fully due systematic uncertainty from the event selection. The reason why long component is so sensitive to the event selection is to be further investigated.

One can see that the nominal threshold is the same for both slow components, regardless of the type of irradiation (gamma rays or alpha particles). Moreover, it is interesting that its value, about 5~Td, is about 1~Td higher than the threshold of excimer EL in Ar (4.1~$\pm$~0.1~Td) \cite{Buzulutskov17,Oliveira11}. This means that the slow components are not related to the excimer EL mechanism.  This also explains why the DarkSide-50 experiment did not observe these unusual slow components: its operational reduced electric field in the EL gap, of 4.6~Td~\cite{Agnes18}, was not enough for this. 

Finally, it was this higher threshold for the appearance of the slow components that made it possible to observe the difference between the fast signal of NBrS EL and slow signal of excimer EL and thus helped to reveal the NBrS effect in proportional EL (see Fig.~19 in \cite{Buzulutskov18}; by chance, the working field there was just at the level of 5.0 Td, i.e. at the verge of appearance of the unusual slow components and therefore was too low for them to interfere).

	\begin{figure}[!t]
	\centering
	\includegraphics[width=1.0\linewidth]{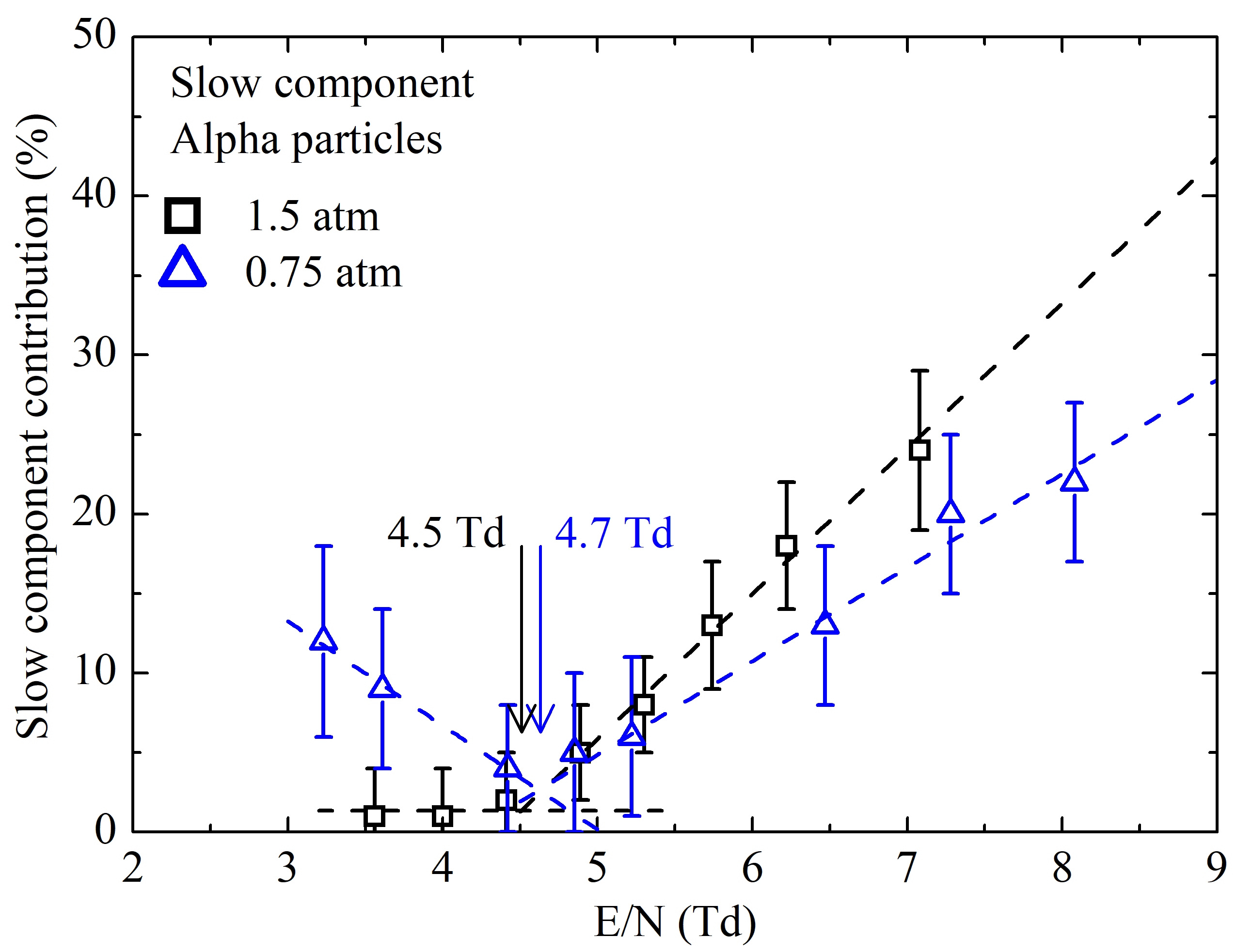}
	\caption{Contribution of the slow component to overall signal as a function of the reduced  electric field, obtained in two-phase Ar detector with $^{238}$Pu alpha particles due to NBrS EL in $\sim$10 mm thick EL gap at two different pressures: 0.75 and 1.5 atm. The arrow points to the nominal threshold of slow component appearance. The PMT and SiPM data are averaged.}
	\vspace{-10pt}
	\label{fig:slow_comp_pressure1}
	\end{figure}

Also, we observed that the slow components threshold did not depend on the gas phase pressure, i.e. on the Ar gas density, as shown in Fig.~\ref{fig:slow_comp_pressure1}: the nominal threshold for slow component appearance was again close to 5~Td and consistent with data at 1~atm pressure within 0.2~Td uncertainty. On the other hand, it is clearly seen that above the threshold, the slow component contribution is larger for higher pressures. 

It should be noted, that the two data points at the lowest electric fields for 0.75~atm are outliers, compared to that of 1.5~atm and 1.0 atm. This is because the absolute electric field in liquid Ar at 0.75~atm was quite low for those points, 1.4 and 1.6~kV/cm, resulting in that the ``usual'' slow component due to electron emission from liquid to gas phase~\cite{Borghesani90,Bondar09} became observable.

The independence of the threshold of the gas density demonstrates that the unusual slow components are not related to the liquid-gas interface. Processes at the liquid-gas interface (known or otherwise) are driven by the absolute electric field, so if the unusual slow components were related to this, the threshold would be constant in the absolute field and not the reduced one. In other words, the unusual slow components appear in the EL gap itself.

	\begin{figure}[!t]
	\centering
	\includegraphics[width=1.0\linewidth]{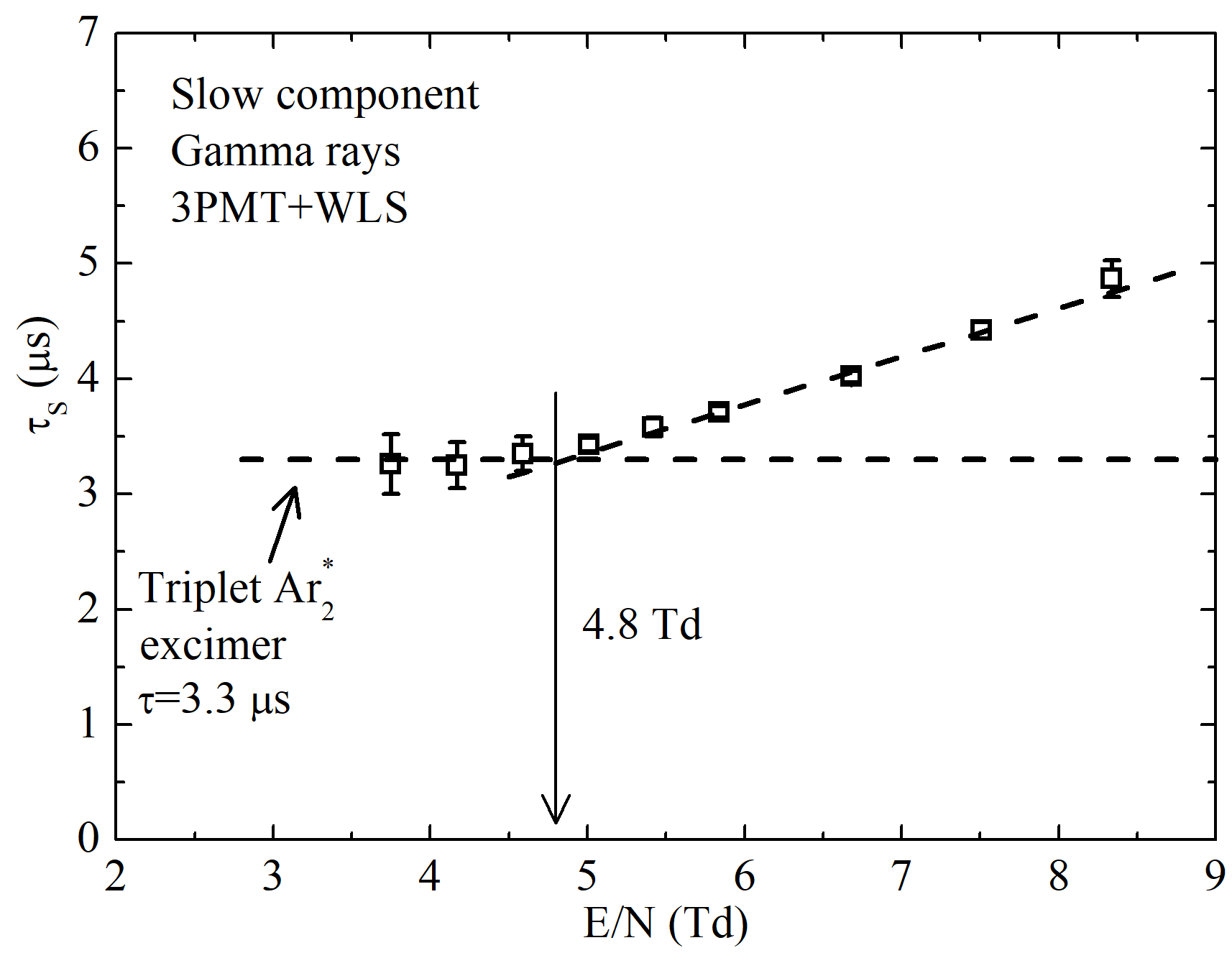}
	\caption{Time constant of the slow component ($\tau_{S}$) as a function of the reduced electric field, obtained in \cite{Bondar20} in two-phase Ar detector with WLS-based readout, i.e. sensitive to excimer EL, with $^{109}$Cd gamma rays in 18~mm thick EL gap at 1.00~atm. The arrow points to the nominal threshold of the time constant increase.}
	\vspace{-10pt}
	\label{fig:slow_comp_WLS}
	\end{figure}

In addition to the results with NBrS EL, the similar threshold behavior of the slow component was observed for excimer EL in time constant dependence on the electric field. Fig.~\ref{fig:slow_comp_WLS} shows the results obtained in our previous work~\cite{Bondar20} in the two-phase detector with WLS-based readout, sensitive in the VUV and thus to excimer EL. In this figure, the triplet slow component of excimer EL with time constant of 3.3~$\mu$s is clearly seen between 4~Td (the threshold for excimer EL in Ar~\cite{Buzulutskov20}) and 5~Td, while above 5~Td the time constant linearly increases with electric field. As discussed in~\cite{Bondar20}, such a specific behavior of the time constant may be explained only if the unusual slow component is added to the charge signal itself above 5~Td, resulting in the appropriate increase of the overall time constant. Note that the nominal threshold for this increase again amounts to about 5~Td.

	\subsection{Correspondence of 5~Td threshold to Ar$^{*}(3p^{5}4p)$ atomic levels and slow to long component ratio}\label{atomic_levels}

	\begin{figure}[t]
	\centering
	\includegraphics[width=1.0\linewidth]{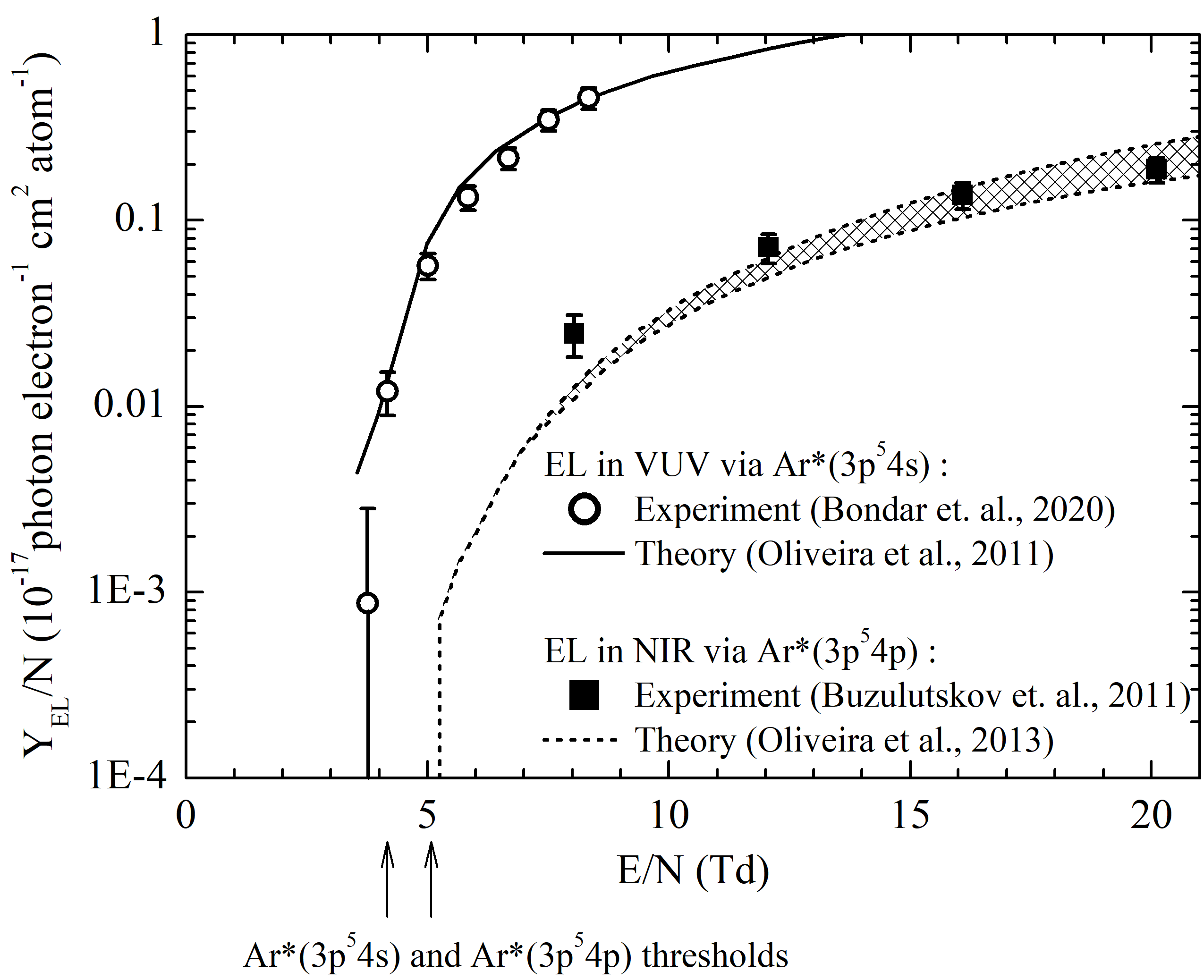}
	\caption{Reduced EL yield in the VUV due to excimer EL, going via Ar$^*(3p^54s)$ excited states, and in the NIR due to atomic EL, going via Ar$^*(3p^54p)$ excited states, measured respectively in~\cite{Bondar20c} and~\cite{Buzulutskov11} (data points). The solid line and the hatched area show the theoretical predictions for those EL mechanisms given in \cite{Oliveira11} and \cite{Oliveira13} respectively. The arrows indicate the thresholds in reduced electric field for those EL mechanisms.}
	\vspace{-10pt}
	\label{fig:Ar_yields}
	\end{figure}

Excimer EL has a certain threshold in reduced electric field, of about 4~Td, defined by the lowest atomic excitation levels Ar$^*(3p^54s)$. This threshold is well seen in Fig.~\ref{fig:Ar_yields} from the EL yield dependence on the electric field, both for theoretical and experimental data. 

Similarly, one can see from this figure that atomic EL has about 1~Td higher threshold compared to excimer EL, at about 5~Td~\cite{Oliveira13}, defined by higher excitation levels Ar$^*(3p^54p)$, which have about 1.5~eV higher energy compared to those of Ar$^*(3p^54s)$.

\begin{figure}[t]
	\centering
	\includegraphics[width=1.0\linewidth]{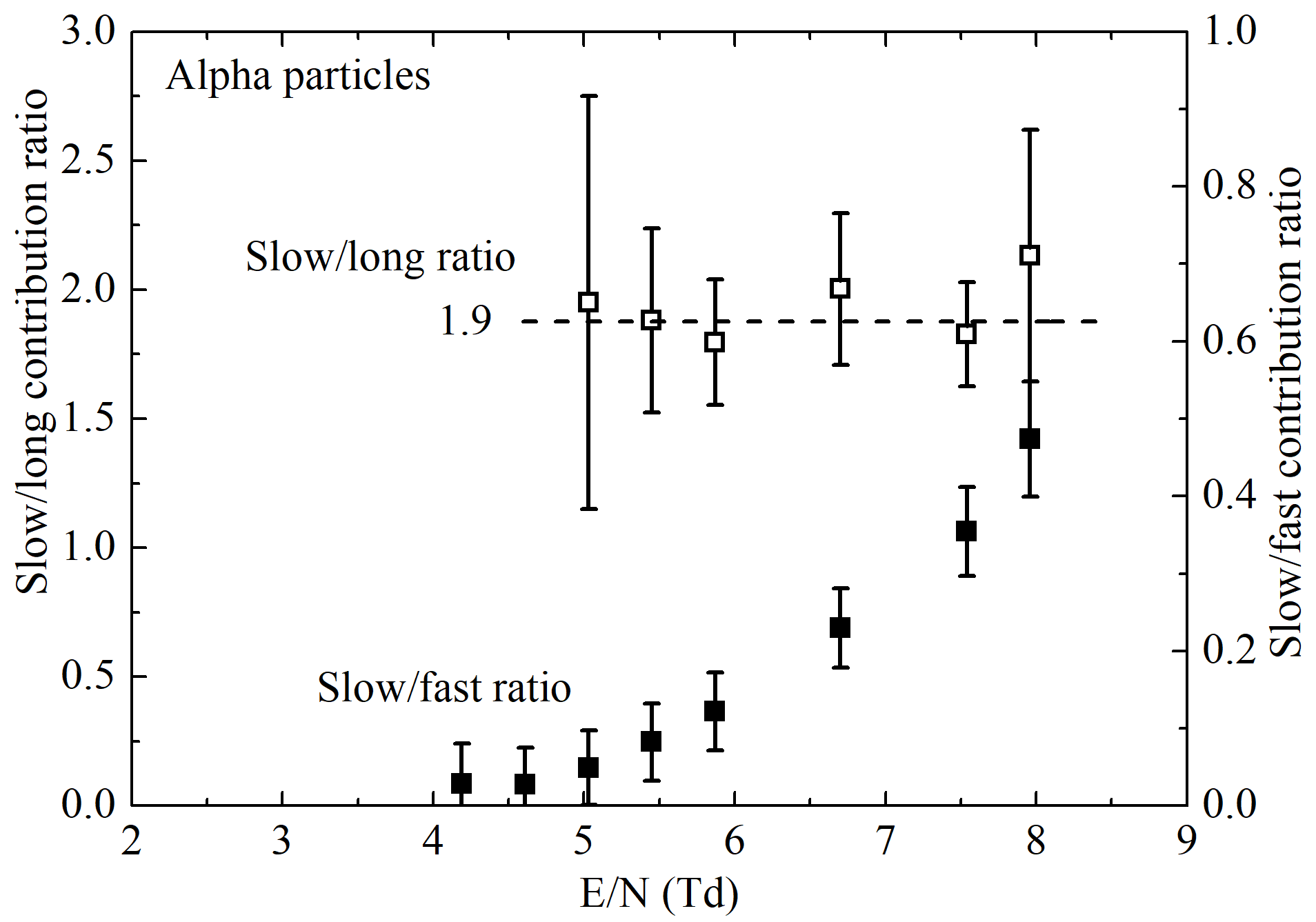}
	\caption{Ratio of slow to long component contribution and that of slow to fast component contribution as a function of the reduced electric field, obtained in two-phase Ar detector with $^{238}$Pu alpha particles due to NBrS EL in 18~mm thick EL gap at 1.00~atm. The PMT and SiPM data are averaged.}
	\vspace{-10pt}
	\label{fig:slow_to_long}
	\end{figure}

Thus one may conclude that the threshold in reduced electric field of 5~Td observed for slow components unambiguously indicates that the higher excited states Ar$^{*}(3p^{5}4p)$ are somehow related to their production mechanism, and that this is not the case for the lower excited states Ar$^{*}(3p^{5}4s)$. This is a remarkable conclusion, since it states that there are some selection rules in the slow component formation mechanism that suppress the inclusion of lower excited states despite their energy favor, and at the same time allow the inclusion of higher excited states.

Also, the fact that the electric field threshold is the same for both slow and long components indicates that these are related to each other and produced by the same mechanism. This conclusion is further confirmed by the electric field independence of the ratio of the slow to long component contribution, in contrast to that of slow to fast component contribution, as shown in Fig.~\ref{fig:slow_to_long}.

	\subsection{Temperature dependence of slow components}\label{results_temperature}

	\begin{figure}[t]
	\centering
	\includegraphics[width=1.0\linewidth]{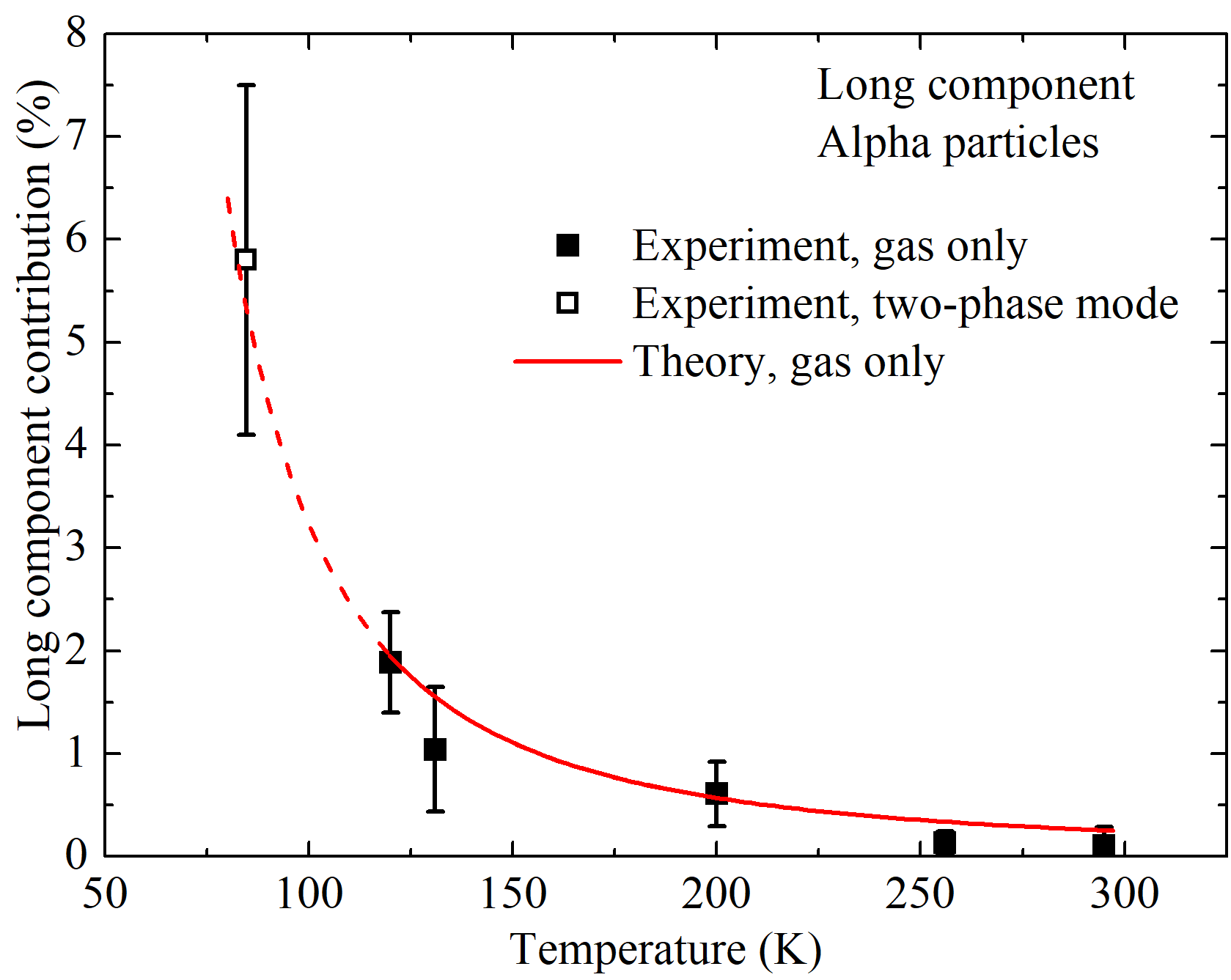}
	\caption{Contribution of the long component to overall signal (at times exceeding 55~$\mu$s with respect to the fast component) as a function of temperature, obtained in single (gas) phase Ar detector with $^{238}$Pu alpha particles for temperatures above 100~K, at fixed gas density (corresponding to 1.5 atm at 295~K) and fixed reduced electric field in the EL gap (6.8~Td). For completeness, the data point at 84.7~K and 0.75~atm obtained in the two-phase mode with 10~mm EL gap is also shown. The solid curve is theoretical fit of single-phase data points by a function proportional to $N_{Ar_2}$ in Eq.~\ref{eq:Ar2}. The curve is extrapolated to the two-phase data point. The PMT and SiPM data are averaged.}
	\vspace{-10pt}
	\label{fig:long_comp_temperature}
	\end{figure}

The temperature dependence of unusual slow components may shed some light on the mechanisms of their formation. To this end, the measurements of S2 pulse shapes were conducted at different temperatures but at the same Ar density. Such measurements were only possible in a single-phase mode, since the two-phase mode dictates that temperature and density are directly linked according to saturated vapor curve. The measurements were conducted at a constant Ar atomic density, of 3.74$\cdot$10$^{19}$~cm$^{-3}$, corresponding to 1.5~atm pressure at room temperature (295 K), and with 5.5~MeV alpha-particle source $^{238}$Pu installed on the cathode as shown in Fig.~\ref{fig:setup_scheme}. The reduced electric field in EL gap was the same for all measurements and amounted to 6.8~Td, with corresponding drift field of 0.24~kV/cm or 0.63~Td.

The mean path length of 5.5~MeV alpha particle at the Ar density used is 3.1~cm which corresponds to 11.7~$\mu$s drift time if the track is collinear to the electric field. For this reason, the tracks orthogonal to the electric field had to be selected in order to localize ionization along z-axis. Otherwise, the resulting EL pulse shapes are complicated by interplay of non-trivial charge distribution and unusual slow components. The track selection was provided using the fast component width, the minimal widths corresponding to orthogonal tracks.

The measured dependence of the long component contribution on Ar gas temperature is shown in Fig.~\ref{fig:long_comp_temperature}. Note that the point at 84.7~K obtained in the two-phase mode and shown in the figure for completeness has the gas-phase density higher by a factor of 1.8 compared to single-phase data. The temperature dependence looks puzzling: the contribution rapidly decreases with temperature and almost disappears above 250~K. Such a dependence is not expected in any of the known scintillation mechanisms. In the next section we try to explain this result.
  
The temperature dependence of the 5~$\mu$s slow component is not shown because of its poor separation from the fast component as well as significant systematical errors due to poor track orthogonality. Nonetheless, preliminary results show that the slow component contribution also decreased with temperature.

\section{Hypothesis of metastable negative argon ions responsible for the formation of slow components}
\label{sec:hypotheis}

All results described above can be successfully explained in the framework of the hypothesis proposed in~\cite{Bondar20}, namely that the unusual slow components appear in the charge signal itself. In this hypothesis, the slow components are produced in the charge signal due to trapping of electrons for some time during their drift in the EL gap, the photon emission before and after trapping being provided by the NBrS EL mechanism. It is proposed that the formation of metastable negative argon ions (Ar$^-$) might be responsible for such trapping. 

Here we suppose that such negative ion states can be produced either in 3-body collisions of drifting electrons with Ar atoms or collisions with Van-der-Waals molecules $\mathrm{Ar}_2(\mathrm{X}\, ^1\Sigma^{+}_{g})$ \cite{Buzulutskov17,Smirnov84,Smirnov96,Stogrin59}, followed by their decay with appropriate time constants:
\begin{eqnarray}
\label{Neg-Ion}
e^- + 2\mathrm{Ar} \rightarrow  \mathrm{Ar}^{-} + \mathrm{Ar}\; , \\
e^- + \mathrm{Ar}_2(\mathrm{X}\, ^1\Sigma^{+}_{g}) \rightarrow  \mathrm{Ar}^{-} + \mathrm{Ar}\; , \\
\mathrm{Ar}^{-} \rightarrow  e^- + \mathrm{Ar}\; . 
\end{eqnarray}

$\mathrm{Ar}_2(\mathrm{X}\, ^1\Sigma^{+}_{g})$ molecule has a rather low binding energy, of 12~meV \cite{Smirnov84,Smirnov96}, which is comparable with the thermal energy at 87~K (10~meV). Therefore, their  content at this temperature, relative to that of the ground states  composed from Ar atoms and Ar$_2$ dimers (see Fig.~\ref{fig:Ar_levels}), is only 2.7\%; it was deduced from~\cite{Stogrin59} using temperature and density dependence of Eq.~\ref{eq:Ar2} taken from \cite{Brahms11}. This content nevertheless might be enough to make this channel dominating, since the reaction of dissociative attachment (Eq. 5) has typically much higher rate than that of three-body collisions (Eq. 4)~\cite{Smirnov74,Massey76}. 

The advantage of this approach is that it can explain the increase of the slow component contribution with electric field: the formation of metastable ions has an energy threshold and the higher the electric field is, the larger the probability of electron having enough energy to become trapped. As discussed in section~\ref{atomic_levels}, both slow and long component have the same threshold in electric field and the ratio of their contributions does not depend on the electric field. As such, unusual slow components have likely the same formation mechanism related to the Ar$^{*}(3p^{5}4p)$ states. This means that either metastable negative argon ions have similar energy level or some intermediate state responsible for their formation is related to Ar$^{*}(3p^{5}4p)$.

In addition, the proposed mechanism of electron trapping predicts the increase of the slow component contribution with gas density observed in experiment (see Fig.~\ref{fig:slow_comp_pressure1}): larger atomic density means higher probability for the electron to form a metastable negative ion during its drift through the EL gap. 

Moreover, in the frame of this approach it is possible to explain the puzzling temperature dependence of the slow components, assuming that they are produced mostly due to electron collisions with Van-der-Waals molecules $\mathrm{Ar}_2(\mathrm{X}\, ^1\Sigma^{+}_{g})$ in reaction (5). In this case, negative ion production rate is proportional to the Ar$_2$ concentration ($N_{Ar_2}$), which is described by the following temperature dependence~\cite{Brahms11}: 
\begin{eqnarray}
\label{eq:Ar2}
N_{Ar_2} =  N_{Ar}^2 \ \lambda_{dB}^3 \  exp(\epsilon_{Ar_2}/k_B T) \; , \\
\nonumber \lambda_{dB} = \sqrt{h^2 /(2 \pi \mu k_B T)} \; ,
\end{eqnarray}
where  $N_{Ar}$ is the Ar atom concentration, $\lambda_{dB}$ is the thermal de Broglie wavelength of the Ar$_2$ molecule of reduced mass $\mu$, $\epsilon_{Ar_2}$ is the Ar$_2$ binding energy, of 12 meV (positive). One can see from Fig.~\ref{fig:long_comp_temperature} that the temperature dependence of Eq.~\ref{eq:Ar2} successfully describes the single-phase data, obtained at a constant gas density and electric field.  

Two possible candidates for the metastable negative ion states were suggested earlier in \cite{Bondar20}: that of Feshbach resonances Ar$^{-}(3p^{5}4s^{2}\ ^{2}P_{1/2,3/2})$ with the energy levels of 11.10 and 11.28~eV and lifetime of about 0.3~ps \cite{Kurokawa2011:Ar_Feshbach} and that of metastable Ar$^{-}(3p^{5}4s4p\:^{4}S)$ state with the energy level of 11.52 eV~and lifetime of about 260~ns~\cite{Bae1985:Ar_ion_state,Ben-Itzhak1988:Ar_ion_state}. For both states the two outer electrons are in excited orbitals, resulting in that their formation should have a threshold in energy and in reduced electric field. The latter should be close to that of excimer EL, i.e. to about 4~Td, due to the close values of the negative ion energy levels to those of the lowest excitation Ar states (Ar$^{*}(3p^{5}4s)$), involved in the excimer formation process (see Fig.\ref{fig:Ar_levels}). In particular, the minimum energy of Ar$^{*}(3p^{5}4s)$ states, 11.55~eV, is very close to 11.52~eV energy of Ar$^{-}(3p^{5}4s4p\:^{4}S)$.

On the other hand, the negative ion states considered above can hardly be responsible for such electron trapping, since their lifetimes and thresholds in electric field are too small compared to observations. Indeed, the simultaneous presence with comparable contributions of the fast component (corresponding to electrons drifting through the EL gap without being trapped) and the slow components indicates that the mean path for electron trapping is of the order of the EL gap thickness, according to Poisson statistics. In other words, two or more electron captures during the drift time are unlikely. This means that the negative Ar ion lifetime should be of the same order as that of the slow components time constants, i.e. of the order of 5 and 50~$\mu$s. This is not the case for Feshbach resonances and metastable Ar$^{-}(3p^{5}4s4p\:^{4}S)$ state. Moreover, the threshold for their formation in reduced electric field of 4~Td does not correspond to the slow component appearance threshold of 5 Td. 

Accordingly, the nature of metastable negative Ar ions responsible for the unusual slow components in EL of two-phase Ar detectors remains unknown.

\section{Conclusions}

In this work, we have for the first time systematically studied the time properties of proportional electroluminescence (EL) in two-phase Ar detectors. Two unusual slow components with time constants of about 4-5 and 50~$\mu$s were observed, which is consistent with our preliminary reports~\cite{Bondar20,Bondar20b}.

Their puzzling property is that their contributions and time constants increase with electric field, which is not expected in any of the known mechanisms of photon and electron emission in two-phase media. In addition, a specific threshold behavior of the slow components was revealed: they emerged at a threshold in reduced electric field of about 5~Td regardless of the gas phase density, which is 1~Td above the onset of standard (excimer) EL, the latter being related to the lower excited atomic states Ar$^{*}(3p^{5}4s)$. 

Accordingly, it is shown that the 5 Td threshold is related to the higher atomic excited states Ar$^{*}(3p^{5}4p)$. This is a remarkable conclusion, since it states that there exist some selection rules in the slow component formation mechanism that suppress the inclusion of the lower excited states despite their energy favor, and at the same time allow the inclusion of the higher excited states.

An unexpected temperature dependence of the 50~$\mu$s component (and presumably that of 5~$\mu$s) was also observed: its contribution decreased with temperature, practically disappearing at room temperature. 

We show that all puzzling properties of the slow components can be successfully explained in the framework of hypothesis that these are produced in the EL gap in the charge signal itself, due to trapping of drifting electrons on metastable negative Ar ions of yet unknown nature with lifetimes close to that of the slow component time constants, i.e. about 5 and 50~$\mu$s. Taking into the account the 5 Td threshold, the formation of these metastable Ar ions is related either to the Ar$^{*}(3p^{5}4p)$ states or some unknown state with similar energy level. In the frame of this approach it is possible to explain the puzzling temperature dependence of the slow components, if one assumes that they are produced mostly at low temperatures due to electron collisions with Van-der-Waals molecules $\mathrm{Ar}_2(\mathrm{X}\, ^1\Sigma^{+}_{g})$. 

The results obtained may have practical applications for the development of two-phase Ar detectors for dark matter searches and low-energy neutrino experiments.

Further studies of the slow component puzzle are in progress in our laboratory.

  \section*{Acknowledgments}

This work was supported in part by Russian Science Foundation (project no. 20-12-00008). The work was done within the R\&D program for the DarkSide-20k experiment.

\bibliographystyle{spphys_modified}       
\bibliography{Manuscript}   

%
%

\end{document}